\documentclass{article}
\usepackage{arxiv}
\usepackage[utf8]{inputenc} % allow utf-8 input
\usepackage[T1]{fontenc}    % use 8-bit T1 fonts
\usepackage{hyperref}       % hyperlinks
\usepackage{url}            % simple URL typesetting
\usepackage{booktabs}       % professional-quality tables
\usepackage{amsfonts}       % blackboard math symbols
\usepackage{amsmath}
\usepackage{nicefrac}       % compact symbols for 1/2, etc.
\usepackage{microtype}      % microtypography
\usepackage{cleveref}       % smart cross-referencing
\usepackage{graphicx}
\usepackage{natbib}
\usepackage{doi}
\usepackage[title]{appendix}

\newcommand{\uncoupled}{ENS-U}
\newcommand{\coupled}{ENS-C}
\newcommand{\hraero}{HRA-ERO}
\newcommand{\lraero}{LRA-ERO}
\newcommand{\hraepo}{HRA-EPO}
\newcommand{\lraepo}{LRA-EPO}

\usepackage{booktabs}
\usepackage{rotating}
\usepackage{siunitx}

\title{The role of the oceans for subseasonal prediction: insights from eddy-permitting and eddy-rich coupled forecast systems}

\author{ \href{https://orcid.org/0000-0002-2958-6637}{\includegraphics[scale=0.06]{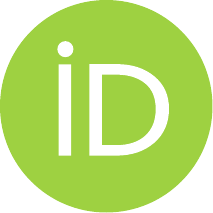}\hspace{1mm}Christopher David Roberts}\\
	ECMWF\\
	Shinfield Park\\
	Reading, United Kingdom\\
	\texttt{chris.roberts@ecmwf.int} \\
	\And
	Sarah Keeley\\
	ECMWF\\
	Shinfield Park\\
	Reading, United Kingdom\\
	\And
	Kristian Mogensen\\
	ECMWF\\
	Shinfield Park\\
	Reading, United Kingdom\\
	\And
	Charles Pelletier\\
	ECMWF\\
    Robert-Schuman-Platz 3\\
    Shinfield Park\\
	Bonn, Germany\\
	\And
	Hao Zuo\\
	ECMWF\\
	Shinfield Park\\
	Reading, United Kingdom\\
}
\renewcommand{\headeright}{PREPRINT}
\renewcommand{\undertitle}{PREPRINT}
\renewcommand{\shorttitle}{The role of the oceans for subseasonal prediction}
\hypersetup{
  colorlinks   = true, %Colours links instead of ugly boxes
  urlcolor     = blue, %Colour for external hyperlinks
  linkcolor    = blue, %Colour of internal links
  citecolor   = red %Colour of citations
}

\begin{document}
\maketitle

%==========================
% A B S T R A C T 
%==========================
\begin{abstract}
The oceans play a fundamental role in Earth's climate system, redistributing heat and influencing global and regional climate variability and predictability across weather and climate timescales. The benefits of ocean-atmosphere coupling for initialised predictions depend on the balance between improvements associated with more realistic air-sea interactions and dynamics, and degradations arising from the development of systematic biases at the coupling interface. Here, we draw on recent developments in modelling and data assimilation at ECMWF to revisit the role of ocean-atmosphere coupling in subseasonal predictions. In particular, we evaluate the impact of ocean-atmosphere coupling in 46-day reforecasts produced with the ECMWF Integrated Forecasting System (IFS) and explore the potential for improvements through increased horizontal resolution and a better representation of the ocean mesoscale. We find that ocean-atmosphere coupling significantly enhances ensemble forecast skill in the tropics, with positive effects increasing at longer lead times. In particular, Madden-Julian Oscillation (MJO) forecasts are substantially improved, with forecast skill extended by approximately 5 days compared to the uncoupled configuration. In contrast, ocean-atmosphere coupling has a more limited impact on the extratropical atmosphere at subseasonal timescales, with marginal impacts on the predictability of major tropospheric and stratospheric circulation indices. Finally, we present selected results from an experimental eddy-rich coupled configuration of the IFS, with a horizontal ocean resolution of approximately 8 km. We find that a better-resolved representation of the ocean mesoscale has a limited impact on atmospheric forecasts at subseasonal lead times, which suggests that many of the known deficiencies of the eddy-permitting reference configuration are mitigated by accurate initialisation. Nevertheless, other factors may be important when considering the case for an eddy-rich ocean in coupled forecasts, including the potential for higher-resolution and more accurate ocean and sea ice forecast products, impacts on coupled data assimilation, and the influence on extreme weather events.

\end{abstract}

%==========================
% I N T R O D U C T I O N 
%==========================
\section{Introduction}
\label{section:intro}

% The role of oceans in Earth's energy budget
The oceans are the principal reservoir of heat in the climate system and interact with the atmosphere through exchanges of energy, moisture, and momentum \citep{trenberth2001estimates, ferrari2009ocean}. These air-sea interactions influence weather and climate variability across all temporal and spatial scales, from planetary-scale overturning circulations to mixed-layer and diurnal processes in the upper ocean \citep{talley2013closure, karlowska2024two}. At global scales, ocean heat uptake regulates the transient climate response to external forcings and plays a dominant role in Earth's energy budget \citep{gregory2008transient, kuhlbrodt2012ocean}. For example, about 90\% of the excess energy associated with top-of-atmosphere (TOA) planetary energy imbalance is absorbed and stored in the oceans \citep{levitus2012world, von2023heat}. The oceans also affect regional climate change and variability through their capacity to store and redistribute heat and salt, with important implications for regional weather and climate patterns \citep{robson2012causes, england2014recent, cassou2018decadal}.

% Physical processes driving coupled predictability at S2D timescales
Internal ocean wave dynamics and basin-scale variations in ocean circulation are also important components of coupled climate variability at seasonal-to-decadal timescales \citep{mcphaden1999genesis, robson2012initialized}. In particular, the El Niño-Southern Oscillation (ENSO) phenomenon arises from coupled interactions between tropical Pacific sea surface temperature (SST), deep convection, the large-scale atmospheric circulation, and equatorial ocean wave dynamics \citep{neelin1998enso}. The atmospheric heating and upper-troposphere circulation anomalies associated with ENSO events also provide a source of atmospheric Rossby waves, which propagate from the tropics to the extratropics thus establishing a clear pathway for the tropical oceans to influence global weather and climate variability \citep{hoskins1981steady, trenberth1998progress, alexander2002atmospheric}. Dynamical representation of the oceans is thus crucial for climate projections and seasonal-to-decadal prediction systems, which rely on the accurate representation of ocean-atmosphere interactions and ocean dynamics to capture predictable signals \citep{griffies1997predictability, stockdale1998global}.

% Physical processes driving coupled predictability at S2S timescales.
Many studies have also highlighted the importance of air-sea interactions for the prediction of atmospheric phenomena at lead times of days to weeks \citep{bender2007operational, woolnough2007role, brassington2015progress, mogensen2017tropical}. Of particular relevance for subseasonal forecasting is the role of ocean feedbacks in the representation of climate modes such as the Madden-Julian Oscillation (MJO), which is the leading mode of intraseasonal variability in the tropics and is characterised by eastward propagating convection and circulation anomalies \citep{madden1972description, demott2015atmosphere, balmaseda2026role}. Through its global teleconnections, the MJO also influences remote weather patterns, including modulation of extratropical weather regimes, tropical cyclone activity, and monsoon systems \citep{hall2001modulation, cassou2008intraseasonal, lin2009observed, liu2022intraseasonal}. Several studies have emphasised that ocean feedbacks can improve the representation of intraseasonal oscillations such as the MJO in global climate models \citep[e.g. ][]{kemball2002simulation, zhang2006simulations, demott2015atmosphere}. However, the impact of ocean feedbacks on the MJO in free-running coupled climate models can be masked by degradations associated with the development of systematic model biases such that the diagnosed impacts of ocean-atmosphere coupling can be sensitive to the simulated mean state \citep{liess2004intraseasonal, sperber2004madden, demott2015atmosphere}. In contrast, initialised predictions provided a cleaner framework to evaluate the impacts of ocean feedbacks on the MJO due to the limited influence of mean state biases and several studies have demonstrated the positive impacts of ocean-atmosphere coupling for subseasonal MJO predictions \citep{woolnough2007role, seo2009evaluation, shelly2014coupled}.

% Eddy-permitting vs eddy-rich ocean models.
State-of-the-art subseasonal prediction systems are commonly coupled to ocean models with an `eddy-permitting' horizontal grid spacing of approximately 25 km \citep{balmaseda2026role}, which is insufficient to fully resolve mesoscale features that scale with the oceanic Rossby radius of deformation (L$_R$) in the mid- and high-latitudes or coastal regions \citep{hallberg2013using, hewitt2017will}. This resolution limitation results in an incomplete or inaccurate representation of important aspects of air-sea interaction and ocean dynamics, which can have significant impacts on the local and remote atmosphere \citep{small2008air, bryan2010frontal, lee2018impact, roberts2021hemispheric}. One manifestation of this problem is that eddy-permitting ocean models often struggle to accurately simulate the location and structure of western boundary currents and their associated sharp gradients in SST \citep{roberts2020time, chassignet2021importance}. In contrast, higher-resolution `eddy-rich' ocean models provide a more realistic representation of many ocean processes, including more realistic ocean eddy activity, improved ocean heat transports, and reduced sea-surface temperature biases \citep{griffies2015impacts, hewitt2017will, hirschi2020atlantic}. However, the computational expense of eddy-rich ocean models has so-far limited their use within operational coupled forecast systems \citep{balmaseda2026role}. Furthermore, the potential benefits of an improved representation of the ocean mesoscale for large-scale atmosphere predictability at subseasonal lead times remains unclear \citep{roberts2022sensitivity, reynolds2025impact}.

% The ECMWF prediction system
In this study, we leverage recent developments in modelling and data assimilation at ECMWF to revisit the role of ocean-atmosphere coupling\footnote{Throughout this study we use `ocean-atmosphere coupling' as shorthand to refer to two-way exchange of information between the ocean, sea ice, ocean wave, and atmosphere components of the ECMWF Integrated Forecasting System (IFS). Uncoupled simulations do not include interactive ocean and sea ice models, but retain two-way coupling between the atmosphere and ocean wave models.} in subseasonal predictions. In particular, we answer the following questions:
\begin{enumerate}
\item What is the role of ocean-atmosphere coupling for large-scale atmospheric predictability at lead times of one to six weeks in the IFS model?
\item Is this assessment sensitive to the specific region, lead time, and atmospheric variable under consideration? 
\item Does increasing ocean model resolution from the `eddy-permitting' to `eddy-rich' regime change these conclusions?
\item Are the diagnosed impacts of an eddy-rich ocean sensitive to atmospheric resolution?
\end{enumerate}
The remainder of this paper is organised as follows: Section \ref{section:methods} describes the methods and datasets used in this study. Section \ref{section:results_coupling} evaluates the impact of ocean-atmosphere coupling in 46-day ensemble reforecasts with the ECMWF IFS model. Section \ref{section:results_resolution} evaluates the impact of increased ocean resolution using selected results from experimental eddy-rich configurations of the IFS, including a scale-dependent evaluation of air-sea interactions and atmospheric variability. Section \ref{section:conclusions} summarises our results and interprets them within the context of previous studies on the role of the oceans for weather and climate modelling. Lastly, this study is a contribution to the special issue \emph{Forecasting in a Changing Climate} and is based on a presentation given by the lead author to the ECMWF 50th anniversary Annual Seminar 2025.

%==========================
% M E T H O D S 
%==========================
\section{Methods}
\label{section:methods}

The European Centre for Medium-Range Weather Forecasts (ECMWF) Integrated Forecasting System (IFS) includes fully-prognostic representations of the atmosphere, ocean, sea ice, land surface, and ocean waves \citep{ifsdoc}. The subseasonal reforecast configurations used in this study are summarised in table \ref{tab:experiments} and described below, with emphasis on changes relative to the IFS cycle 47r3-based configurations described by \citet{roberts2023euro}. The eddy-rich and associated eddy-permitting reference experiments presented in section \ref{section:results_resolution} are conducted with IFS cycle 48r1, whereas the coupling sensitivity experiments used in section \ref{section:results_coupling} are based on the more recent IFS cycle 49r2. These experiments were conducted at different times and the high computational cost of eddy-rich simulations precluded rerunning them with a common IFS cycle. Similarly, the reforecast periods differ between sections due to the availability of suitable ocean initial conditions for each configuration. Nevertheless, the absolute forecast performance of the two IFS cycles is sufficiently similar that scientific insights from one configuration remain relevant for the other, and our main conclusions are robust to these differences, which are highlighted below.

\subsection{Coupled subseasonal ensemble reforecast (\coupled{})}
% EXPID=impq

Our reference coupled ensemble reforecasts are conducted with IFS cycle 49r2. Although this cycle is not operational for medium-range or subseasonal forecasting at ECMWF, it serves as the foundation for the next ECMWF reanalysis and seasonal forecasting systems. The most relevant model changes for subseasonal forecasting between cycles 47r3 and 49r2 include the following: 

\begin{itemize}
\item \textbf{Cycle 48r1}: The introduction of a multi-layer snow scheme; revised surface boundary conditions; optimisations to parameterised gravity wave drag; a new hybrid-linear ozone scheme; and several other changes, including updates to parameterised microphysical processes \citep{arduini2019impact, kanehama2022evaluation, lang2023ifs}.

\item \textbf{Cycle 49r1}: Activation of the stochastically perturbed parameterisations (SPP) scheme for atmospheric model uncertainty; revisions to the ocean wave model, including an increase of horizontal resolution to match the native IFS atmosphere grid; and revisions to the land surface model, including a new urban canopy model \citep{ollinaho2017towards, leutbecher2024improving, roberts2024ifs}.

\item \textbf{Cycle 49r2}: Implementation of new ocean and sea ice configurations and associated (re)analyses based on version 4.0.6 of the Nucleus for European Modelling of the Ocean (NEMO) ocean model and the Sea Ice modelling Integrated Initiative (SI$^3$) sea ice model \citep{madec2019nemo, vancoppenolle2023si3,keeley2024introduction}, which will become operational in IFS cycle 50r1 \citep{polichtchouk2025upgrade}. 
\end{itemize}

The most relevant developments for this study are the changes to the ocean and sea ice models, which include revisions to the ocean turbulent mixing following \citet{storkey2018uk}; activation of a nonlinear free surface and z-star vertical coordinate to provide local conservation of volume and tracers; and the introduction of a multi-category sea ice model, which includes prognostic salinity and a parameterised representation of surface melt-ponds \citep{keeley2024introduction}. Other important changes include more consistent coupling of sea ice and snow depth information between the sea ice and atmosphere models; deactivation of partial coupling in the extratropics, ensuring full ocean-atmosphere coupling from the first time step; and the use of reduced-precision arithmetic across all model components--including ocean and sea ice--which improves computational efficiency without compromising accuracy \citep[e.g. ][]{lang2021more}.

These physical modelling updates are accompanied by new ocean and sea ice initial conditions from the 6th generation ECMWF ocean and sea ice ensemble reanalysis system \citep[ORAS6; ][]{zuo2024ecmwf}. In addition to adopting the revised physical models, ORAS6 introduces several other developments, including variational assimilation of SST observations, hourly surface forcing from ERA5 \citep{hersbach2020era5}, and a larger ensemble of ten perturbed members and one control member. Together, these changes to data assimilation methods and model formulation yield important improvements over ORAS5 \citep{zuo2019ecmwf}, including a better representation of the diurnal cycle in SST, reduced near-surface biases in the Gulf Stream region, and substantially improved near-surface temperature and sea level variability throughout the extratropics \citep{zuo2024ecmwf}.

The coupled reference ensemble reforecast used in this study (\coupled) consists of 10 perturbed members, each integrated for 46-days, starting on the first of each month between 2006 and 2023 for a total of 216 start dates. The atmospheric model is configured to use the cubic octahedral reduced Gaussian grid with 137 vertical levels and a horizontal resolution of Tco319 ($\Delta$x $\approx$ 35 km). The atmosphere is coupled hourly to an eddy-permitting NEMO4-SI$^3$ configuration, which uses 75 vertical levels in the ocean and the eORCA025 grid ($\Delta$x $\approx$ 25 km). Atmospheric fields are initialised from the ERA5 reanalysis and ocean/sea-ice fields are initialised using the ORAS6 ensemble reanalysis. Ensemble spread in coupled forecasts is generated through a combination of atmosphere, ocean, and sea ice initial condition perturbations and stochastic parameterisations in the atmospheric model \citep{leutbecher2017stochastic, lock2019treatment, leutbecher2024improving,zuo2024ecmwf}.

\subsection{Uncoupled subseasonal ensemble reforecasts (\uncoupled{})}
% EXPID=imq1

\begin{figure}[!htbp]
    \includegraphics[width=14cm]{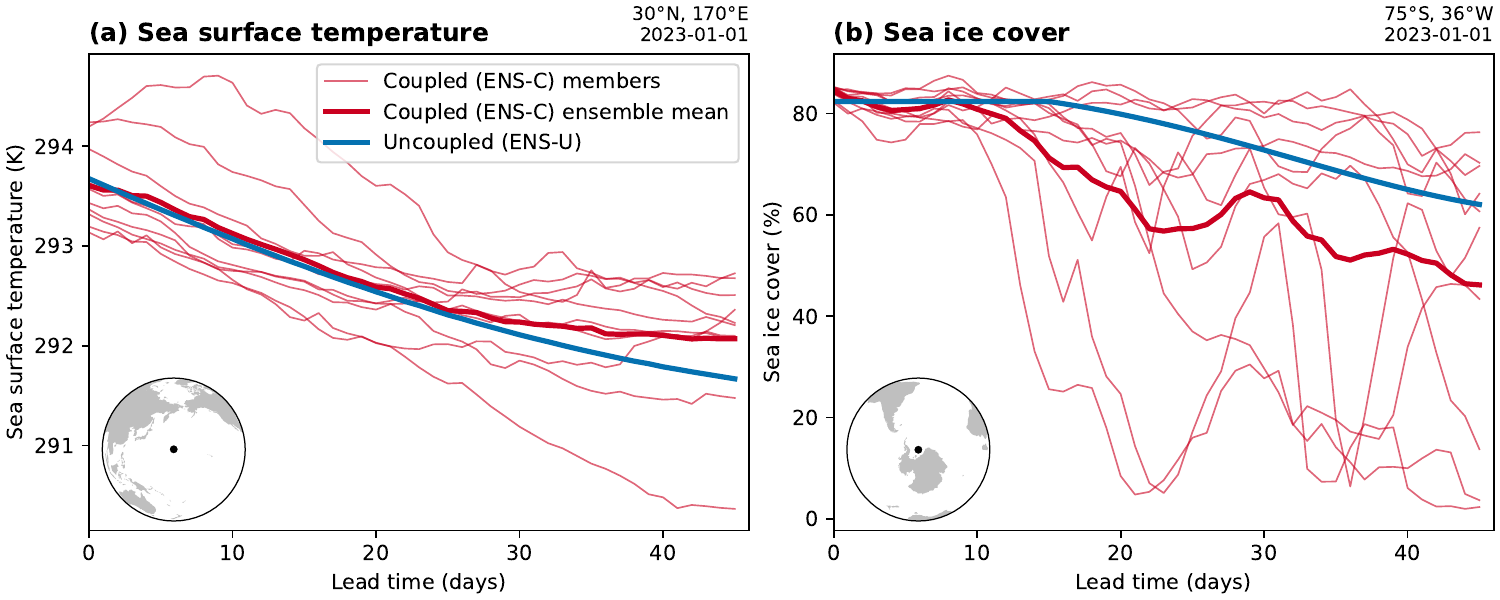}
    \centering
    \caption{(a) Example of sea surface temperatures vs lead time in coupled (\coupled{}) and uncoupled (\uncoupled{}) ensemble forecasts initialised on January 1st 2023, at the grid point nearest to 30$^{\circ}$N, 170$^{\circ}$E in the North Pacific. Thin red lines show the ten perturbed members of \coupled{} and the thick red line shows their ensemble mean. Each member in \coupled{} is initialised with a different member from the ORAS6 ensemble reanalysis \citep{zuo2024ecmwf}, whereas SST boundary conditions for \uncoupled{} are shared across all members and are shown as a single thick blue line. (b) As for SSTs, but for sea ice cover at the grid point nearest to 75$^{\circ}$S, 36$^{\circ}$W in the Weddell Sea. Inset globes mark the location of the selected grid point in each panel.}
    \label{fig:coupled_vs_uncoupled_ssts}
\end{figure}

To evaluate the impact of ocean-atmosphere coupling, we run subseasonal ensemble reforecasts with ocean and sea ice boundary conditions derived from observed values at initialisation time (\uncoupled{}). In these experiments, SSTs are specified using daily values from the Operational Sea Surface Temperature and Ice Analysis \citep{donlon2012operational} that are persisted as anomalies on top of a daily mean SST climatology (1979-2001) derived from the ERA40 reanalysis \citep{uppala2005era, jung2008scale}. Sea ice cover is specified using persistence of Ocean and Sea Ice Satellite Application Facility (OSI SAF) initial values for the first 15 days \citep{eastwood2014algorithm}, linear blending between persistence and the ERA40 climatology for days 16-45, and climatology thereafter. The difference between coupled and uncoupled surface boundary conditions for an example start date is illustrated in figure \ref{fig:coupled_vs_uncoupled_ssts}. This approach follows the specification of SST and sea ice boundary conditions in ECMWF operational deterministic forecasts prior to the activation of ocean-atmosphere coupling on June 5th 2018 \citep{mogensen2017tropical,keeley2018dynamic}. This persistence-climatology approach represents a realistic operational alternative to a fully coupled ocean model, since observed SSTs beyond the initialisation date are not available in real-time forecast applications. Other than the specified ocean and sea ice boundary conditions, these reforecasts are configured identically to the \coupled{} experiments.

\subsection{Eddy-permitting subseasonal reforecasts (\lraepo{} and \hraepo{})}
% Tco319 EXPID=i7au
% Tco1279 EXPID=ii61 
In addition to the coupled and uncoupled ensemble reforecasts described above, we also present evidence from single-member reforecasts run with different ocean and atmosphere resolutions. Ensemble forecasts were not run for these configurations due to the combination of (i) their high computational cost and (ii) the absence of suitable ocean initial conditions to initialise an eddy-rich ensemble. The eddy-permitting reforecasts are based on IFS cycle 48r1 atmosphere run at Tco319 ($\Delta$x $\approx$ 35 km) and Tco1279 ($\Delta$x $\approx$ 9 km) combined with the eORCA025 NEMO4-SI$^3$ ocean and sea ice configurations ($\Delta$x $\approx$ 25 km) described above. We refer to these configurations as \emph{lower-resolution atmosphere eddy-permitting ocean} (\lraepo{}) and \emph{higher-resolution atmosphere eddy-permitting ocean} (\hraepo{}), respectively. Each reforecast consists of a single unperturbed member run for 32 days starting on the first of each month between 1995 and 2016, for a total of 264 start dates. Atmospheric fields are initialised from the ERA5 reanalysis \citep{hersbach2020era5} and ocean/sea-ice initial conditions are derived from ERA5-forced NEMO4-SI$^3$ simulations that are tightly constrained to follow daily values from the GLORYS12v1 eddy-rich global ocean reanalysis \citep[][]{jean2021copernicus}. Further details on the generation of ocean and sea ice initial conditions are provided in Appendix A. 

% -------------------------
% S T A R T   T A B L E 
% -------------------------
\begin{table}[!htbp]
\centering
\caption{Summary of subseasonal reforecast configurations used in this study.}
\label{tab:experiments}
\renewcommand{\arraystretch}{1.3}
\footnotesize
\begin{tabular}{lcccccc}
\hline\
\textbf{Experiment} & \textbf{Atmosphere} & \textbf{Ocean/ice} & \textbf{Members} & \textbf{Length (days)} & \textbf{Period} & \textbf{Initial conditions} \\
\hline
\coupled{}   & IFS 49r2 (${\sim}35$ km) & NEMO4-SI$^3$ (${\sim}25$ km) & 10 & 46 & 2006--2023 & ERA5/ORAS6 \\
\uncoupled{} & IFS 49r2 (${\sim}35$ km) & Persistence/climatology $^{a}$ & 10 & 46 & 2006--2023 & ERA5 \\
\lraepo{}    & IFS 48r1 (${\sim}35$ km) & NEMO4-SI$^3$ (${\sim}25$ km)  & 1  & 32 & 1995--2016 & ERA5/GLORYS12v1$^{b}$ \\
\hraepo{}    & IFS 48r1 (${\sim}9$ km) & NEMO4-SI$^3$ (${\sim}25$ km) & 1  & 32 & 1995--2016 & ERA5/GLORYS12v1$^{b}$ \\
\lraero{}    & IFS 48r1 (${\sim}35$ km) & NEMO4-SI$^3$ (${\sim}8$ km)  & 1  & 32 & 1995--2016 & ERA5/GLORYS12v1$^{b}$ \\
\hraero{}    & IFS 48r1 (${\sim}9$ km) & NEMO4-SI$^3$ (${\sim}8$ km)  & 1  & 32 & 1995--2016 & ERA5/GLORYS12v1$^{b}$ \\
\hline\hline
\multicolumn{7}{l}{$^{a}$ SST and sea ice boundary conditions derived from observations using a combination of (anomaly) persistence and climatology.} \\
\multicolumn{7}{l}{$^{b}$ Ocean and sea ice initial conditions derived from GLORYS12v1 as described in Appendix A.} \\
\end{tabular}
\end{table}
% -------------------------
% E N D   T A B L E
% -------------------------

\subsection{Eddy-rich subseasonal reforecasts (\lraero{} and \hraero{})}
% Tco319 EXPID=i7u3
% Tco1279 EXPID=1162
Eddy-rich subseasonal reforecasts are based on the IFS cycle 48r1 atmosphere run at Tco319 ($\Delta$x $\approx$ 35 km) and Tco1279 ($\Delta$x $\approx$ 9 km) combined with the eORCA12 NEMO4-SI$^3$ ocean and sea ice configurations ($\Delta$x $\approx$ 8 km). We refer to these configurations as \emph{lower-resolution atmosphere eddy-rich ocean} (\lraero{}) and \emph{higher-resolution atmosphere eddy-rich ocean} (\hraero{}), respectively. Other than the increased ocean resolution, these configurations are identical to the \lraepo{} and \hraepo{} configurations described above, with ocean and sea ice initial conditions derived from GLORYS12v1 as described in Appendix A.

\subsection{Atmospheric circulation indices}
To complement our analysis of weekly mean gridded fields, we also evaluate the impact of ocean-atmosphere coupling on four daily mean indices that measure different aspects of the large-scale tropospheric and stratospheric circulation that are important at subseasonal timescales. Specifically, we consider the Madden-Julian oscillation (MJO), the North Atlantic Oscillation (NAO), the Pacific-North American pattern (PNA), and the Northern Hemisphere Stratospheric Polar Vortex (PVORTEX). A detailed description of the calculation of these indices from ERA5 and ECMWF reforecast data is provided by \citet{roberts2025ensemble}, which we summarise below.

MJO predictability is evaluated using the daily mean real-time multivariate MJO (RMM) index following \citet{wheeler2004all} and \citet{gottschalck2010framework}. The two components of the bivariate RMM index (RMM1 and RMM2) are used to define MJO amplitude and phase as $\sqrt{\textnormal{RMM1}^2 + \textnormal{RMM2}^2}$ and $\textnormal{arctan2}(\textnormal{RMM2}, \textnormal{RMM1})$, respectively. Phase numbers correspond to the different sectors of MJO phase diagram and are indicative of MJO activity over the Indian Ocean (phases 2 and 3), maritime continent (phases 4 and 5), western Pacific Ocean (phases 6 and 7), and the Atlantic Ocean/Africa (phases 8 and 1). NAO indices are derived by projecting 500 hPa geopotential height anomalies onto a loading pattern defined as the first empirical orthogonal function (EOF) for the region bounded by 20$^{\circ}$N-80$^{\circ}$N and 90$^{\circ}$W-40$^{\circ}$E. PNA indices are calculated the same way, but for the region 10$^{\circ}$N-80$^{\circ}$N and 150$^{\circ}$E-300$^{\circ}$E. The PVORTEX index is calculated as the normalized zonal mean of 50 hPa zonal wind anomalies at 60$^{\circ}$N.

\subsection{Reanalysis and satellite data}
Coupled and uncoupled ensemble forecasts are evaluated against the ERA5 reanalysis \citep{hersbach2020era5} on a reduced-resolution $2.5^{\circ}\times2.5^{\circ}$ latitude-longitude grid, which reflects our focus on the predictability of the large-scale atmospheric circulation. Eddy-permitting (\lraepo{} and \hraepo{}) and eddy-rich (\lraero{} and \hraero{}) reforecasts are verified using a higher-resolution $0.5^{\circ}\times0.5^{\circ}$ latitude-longitude grid against a combination of ERA5 for atmospheric fields and reprocessed satellite products for ocean and sea ice fields. During model integrations, all ocean and atmosphere fields are archived on the native atmospheric grid before interpolation to a common verification grid using either conservative (ocean and surface fields) or linear (atmosphere pressure levels) methods. For this reason, our evaluation of ocean and sea ice fields should be considered an evaluation of the fields as seen by the atmosphere, and is thus sometimes sensitive to the resolution of the intermediate atmospheric grid and choice of verification grid. We highlight these sensitivities in our results where relevant.

For the high-resolution evaluation of eddy-permitting and eddy-rich experiments, SSTs are compared with the satellite-based European Space Agency-Climate Change Initiative/Copernicus Climate Change Service (ESA-CCI/C3S) reprocessed product \citep{merchant2019satellite, good2020current}. Sea ice cover is evaluated using OSI-430/OSI-450 data distributed alongside the ESA-CCI/C3S SSTs, which are originally provided by the Ocean and Sea Ice Satellite Application Facility \citep[OSI-SAF; ][]{lavergne2019version}. Sea surface heights are evaluated using the reprocessed analysis distributed by the Copernicus Marine Service, which is produced by the Data Unification and Altimeter Combination System (DUACS) multimission altimeter data processing system \citep{sealevel_product}. For consistent comparison between models and satellite observations, we remove the global mean from both simulated and observed sea-surface height fields. We follow \citet{roberts2016drivers} and refer to this derived property as \emph{dynamic sea level}, as it can be directly related to changes in ocean circulation and does not include contributions from global thermosteric sea-level rise or changes to the total mass of the ocean. Near-surface wind speeds over the ocean are evaluated using the Copernicus Marine Service global L4 wind product \citep{wind_product}, which consists of hourly ERA5 10 m stress-equivalent winds that are corrected for differences relative to collocated scatterometer observations. Daily mean wind speeds are derived from 12-hourly values for consistency with the reforecasts. 

\subsection{Verification methods}
The impacts of coupling and changes in ocean resolution are assessed by separately evaluating the impact on the climatological mean state and forecast anomalies calculated following the `by-member--other-years' method described in \citet{roberts2025unbiased}. Impacts on the mean state are assessed by computing relative changes in the mean absolute bias (MAB) using a mean absolute bias score (MABS), defined as:
\begin{equation}
\mathrm{MABS} = 1 - \frac{\mathrm{MAB}_{\mathrm{A}}}{\mathrm{MAB}_{\mathrm{B}}},
\label{eq:mabs}
\end{equation}
where
\begin{equation}
\mathrm{MAB} = 
\frac{
    \sum_{i} w_i \sum_{m} \left| \langle F \rangle_{i,m} - \langle O \rangle_{i,m} \right|
}{
    \sum_{i} w_i
},
\end{equation}
and $\langle F \rangle_{i,m}$ and $\langle O \rangle_{i,m}$ represent climatologies for a specific lead time in forecasts and observations, respectively, as a function of grid-point $i$ and the month of the forecast start date $m$, and $w_i$ is a weight proportional to grid-cell area. A positive value of MABS indicates that absolute biases aggregated across all locations and start dates are reduced in experiment A compared to experiment B. The aggregation of absolute biases calculated separately for each month means that MABS is also sensitive to improvements in the amplitude of the seasonal cycle that occur without an associated change to the annual mean climatology.

The impacts on deterministic and probabilistic skill of anomaly forecasts are evaluated using anomaly correlations, root mean square error (RMSE), and the fair version of the continuous ranked probability score \citep[fCRPS; ][]{ferro2008effect, ferro2014fair, leutbecher2019ensemble}. Skill scores are computed from area-weighted scores averaged over all forecast cases for a given lead time. The area-weighted regional fCRPS is calculated as
\begin{equation}
\overline{\mathrm{fCRPS}}_{\mathrm{region}} = \frac{\sum_{i} w_{i} \cdot \overline{\mathrm{fCRPS}}_{i}}{\sum_{i} w_{i}},
\end{equation}
where the overbar denotes an average over all forecast cases for a given lead time. The associated skill score is given by 
\begin{equation}
\mathrm{fCRPSS}_{\mathrm{region}} = 1 - \frac{ \overline{\mathrm{fCRPS}}_{\mathrm{region}} }{ \overline{\mathrm{CRPS0}}_{\mathrm{region}} },
\end{equation}
where $\overline{\mathrm{CRPS0}}_{\mathrm{region}}$ represents the unadjusted CRPS of reference forecasts constructed from the climatological distribution of observed anomalies (excluding the forecast start date), averaged in the same manner. For deterministic forecasts, we also calculate area-weighted RMSE as
\begin{equation}
\mathrm{RMSE}_{\mathrm{region}} = \sqrt{ \frac{\sum_{i} w_{i} \cdot \mathrm{MSE}_{i}}{\sum_{i} w_{i}}},
\end{equation}
where MSE$_{i}$ is the mean squared error for a specific location and lead time. Sampling uncertainty is evaluated using a bootstrap resampling approach where scores (and associated differences) are calculated 500 times using randomly selected (with replacement) start dates as described in \citet{roberts2022sensitivity}. Score differences are deemed statistically robust if the 2.5$^{th}$ and 97.5$^{th}$ percentiles of the bootstrap distribution have the same sign.

\subsection{Scale-dependent diagnostics}
\label{section:methods_scale}
\subsubsection{SST-wind relationships}
To evaluate air-sea interactions at the ocean mesoscale, we follow \citet{bryan2010frontal} and diagnose scale-dependent local relationships between 10 m wind speed and SST anomalies. We use weekly mean anomalies of 10 m wind speed (calculated from 12-hourly instantaneous values) and SST (calculated from daily means) on a common $0.25^{\circ}\times0.25^{\circ}$ latitude-longitude grid relative to start date and lead time dependent climatologies that are calculated consistently for each reforecast configuration and observational dataset. Low-passed fields are obtained by smoothing each weekly mean anomaly with a 6$^{\circ}$ latitude by 18$^{\circ}$ longitude boxcar filter, which is similar to the loess filter with half-power points at 10$^{\circ}$ latitude and 30$^{\circ}$ longitude used by \citet{bryan2010frontal}. High-passed anomalies are derived as the difference between the original anomalies and low-passed anomalies. 

We calculate correlations and least-squares regression slopes of wind speed on SST for each grid point using all forecasts initialised in a given season for a fixed lead time. The CMEMS  observational wind product is only available from 2007 so this analysis is restricted to forecasts initialised between 2008 and 2016. Observation-based estimates are calculated in the same way using ESA-CCI SSTs and Copernicus Marine Service winds (see section \ref{section:methods}). Regional mean regression slopes are calculated as cosine-latitude weighted averages over the Gulf Stream (37$^{\circ}$N-45$^{\circ}$N, 75$^{\circ}$W-40$^{\circ}$W), the Southern Ocean (35$^{\circ}$S-55$^{\circ}$S), and the Kuroshio (35$^{\circ}$N-42$^{\circ}$N, 140$^{\circ}$E-170$^{\circ}$E), where the boxes for the Gulf Stream and Kuroshio are chosen to enclose the regions of positive high-passed correlations in both seasons. Results are only shown equatorward of 60$^{\circ}$ latitude to avoid regions affected by sea ice. Our approach differs from \citet{bryan2010frontal} in the use of weekly rather than monthly means, but is otherwise designed to isolate the same frontal-scale relationships.

\subsubsection{Atmospheric spectra}
To evaluate whether increased ocean resolution modifies the distribution of atmospheric variability across spatial scales, we calculate spectra of temperature variance and horizontal kinetic energy at 850 and 200 hPa in our eddy-permitting and eddy-rich reforecasts. Spectra are computed from the exact spherical harmonic coefficients of the IFS model fields. Temperature variance spectra are expressed as a function of total wavenumber $n$ and normalised such that their sum over $n \geq 1$ is equal to the area-weighted global variance. Kinetic energy spectra (per unit mass) are calculated from the spherical harmonic coefficients of relative vorticity and divergence, such that their sum is equal to the global mean kinetic energy. Each total wavenumber is associated with an equivalent wavelength of $2\pi a / n$, where $a$ is the radius of the Earth. Spectra are calculated from instantaneous fields at 00 UTC on day 32 of all forecasts initialised on a specific reference date (e.g. January 1st) and then averaged over all start dates. To quantify differences between eddy-rich and eddy-permitting configurations, spectra are summed over a range of total wavenumbers for each start date. Differences are then expressed as a percentage of the eddy-permitting value with 95\% confidence intervals estimated by bootstrap resampling over start dates.

%==========================
% R E S U L T S
%==========================
\section{Impacts of ocean-atmosphere coupling in ECMWF S2S forecasts}
\label{section:results_coupling}
\subsection{Impacts on the mean state}
\begin{figure}[!htbp]
    \includegraphics[width=14cm]{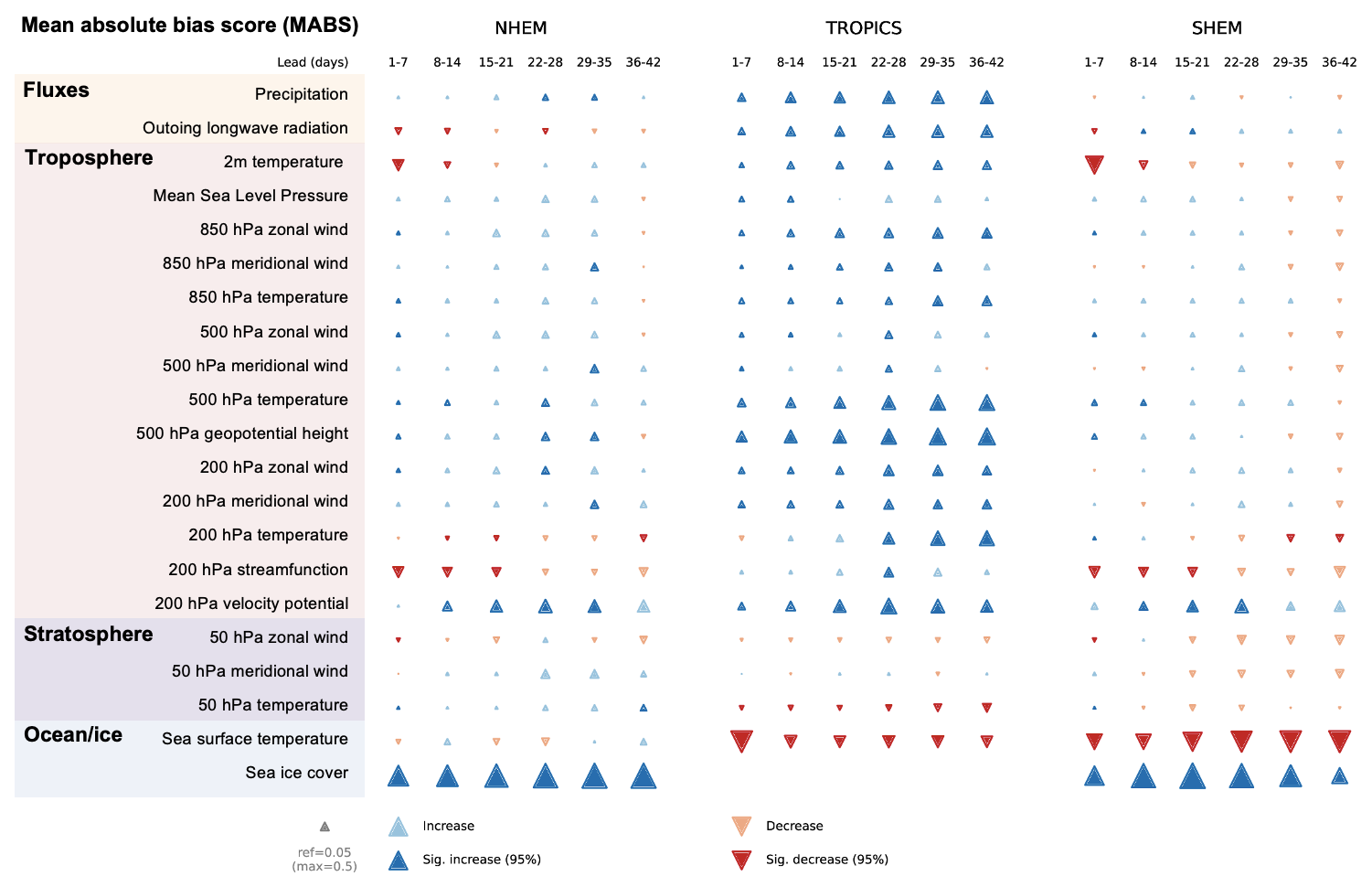}
    \centering
    \caption{A score card summarizing the impact of ocean-atmosphere coupling on weekly mean climatologies for different lead times and the following regions: NHEM (30$^{\circ}$N-90$^{\circ}$N), TROPICS (30$^{\circ}$S-30$^{\circ}$N), and SHEM (30$^{\circ}$S-90$^{\circ}$S). Blue triangles indicate positive values of MABS (equation \ref{eq:mabs}), which correspond to improvements in \coupled{} compared to \uncoupled{} relative to the ERA5 reference data. Red triangles indicate negative values of MABS, which corresponds to degradations in \coupled{} compared to \uncoupled{}. Symbol areas are proportional to the magnitude of MABS and darker triangles indicate that bootstrap-derived 95\% confidence intervals do not intersect zero. The area of the grey reference triangle corresponds to $\mathrm{MABS}=0.05$.}
    \label{fig:mabs}
\end{figure}

To begin, we summarise the impacts of ocean-atmosphere coupling on the mean state of ECMWF subseasonal forecasts for a range of parameters, regions, and lead times (figure \ref{fig:mabs}). In these experiments, ocean-atmosphere coupling has a small but negative impact on the climatological mean state of SSTs as diagnosed from the statistically robust differences in MABS for SST across all lead times in the tropics and southern hemisphere. However, these large relative changes correspond to rather small differences in absolute terms. For example, the week 1 mean absolute bias of SST over the tropics increases from 0.08 K in ENS-U to 0.10 K in ENS-C. In contrast, coupling improves the climatological mean sea ice cover, which likely reflects deficiencies in the crude persistence-climatology approximation in ENS-U compared to the combination of coupled forecasts and high quality sea ice analysis initial conditions in ENS-C. 

In general, the tropical atmosphere mean state is more sensitive to ocean-atmosphere coupling than the extratropical atmospheric mean state. However, we identify several atmospheric fields where the climatological mean is slightly degraded during the first few weeks of ENS-C forecasts, including outgoing longwave radiation, two-metre temperature, and 200 hPa streamfunctions in the extratropics. There is also some evidence for very small relative degradations to climatological 50 hPa temperatures in the tropics. Nevertheless, the small degradations to SSTs in the tropics and southern hemisphere do not generally correspond to negative impacts on the atmospheric mean state. On the contrary, the mean state of the tropical atmosphere is generally improved across a wide range of parameters and lead times (figure \ref{fig:mabs}), which is indicative of a non-linear rectified impact of ocean-atmosphere coupling on the atmospheric mean state mediated by improved air-sea interactions and coupled ocean-atmosphere variability. This interpretation is supported by our evaluation of weekly forecast anomalies in the following section.

\subsection{Impacts on weekly mean forecast anomalies}
\begin{figure}[!htbp]
    \includegraphics[width=14cm]{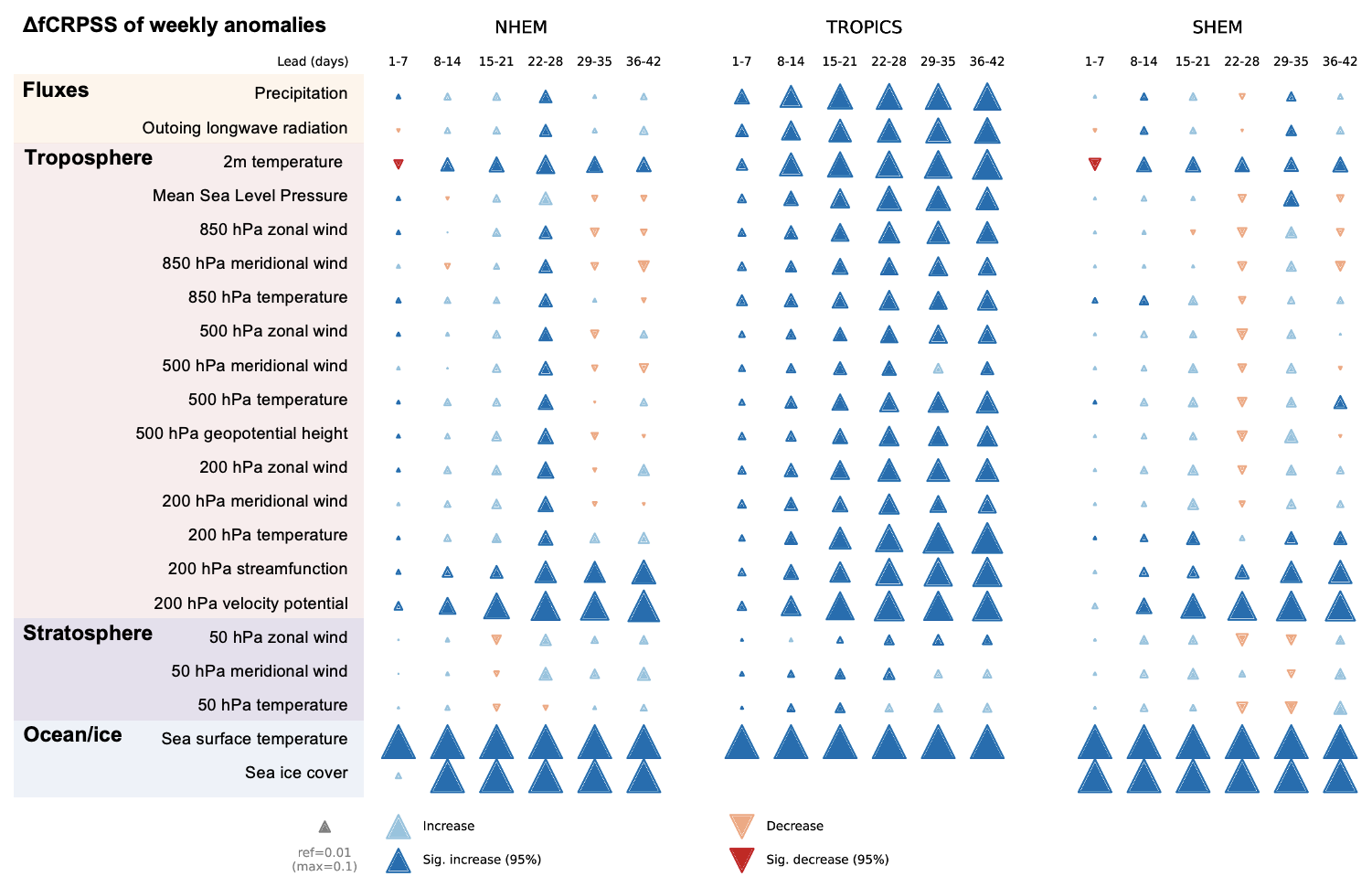}
    \centering
    \caption{As figure \ref{fig:mabs}, but for differences in fCRPSS calculated using weekly mean anomalies. Blue triangles indicate positive values of $\Delta$fCRPSS, which correspond to improved anomaly-based forecast skill in \coupled{} compared to \uncoupled{}. Red triangles indicate negative values of $\Delta$fCRPSS, which corresponds to degradations in anomaly-based forecast skill in \coupled{} compared to \uncoupled{}. Symbol areas are proportional to the magnitude of $\Delta$fCRPSS and darker triangles indicate that bootstrap-derived 95\% confidence intervals do not intersect zero. The area of the grey reference triangle corresponds to $\Delta\mathrm{fCRPSS}=0.01$.}
    \label{fig:delta_fcrpss}
\end{figure}

Ocean-atmosphere coupling has unequivocally positive impacts on weekly mean SST and sea ice anomaly forecasts across all subseasonal lead times (figure \ref{fig:delta_fcrpss}). These large improvements in \coupled{} mostly reflect the crude persistence-climatology approximation used in \uncoupled{}, which diverges from the true ocean state at extended lead times. Coupling also improves anomaly-based forecast skill for all evaluated fields in the tropical atmosphere, which reflects the strong coupling between SST anomalies, turbulent air-sea fluxes, atmospheric convection, and the large-scale circulation in the tropical troposphere \citep{demott2015atmosphere}. These positive impacts on forecast skill are most pronounced in the troposphere, where they grow with lead time, but more limited in the stratosphere (figure \ref{fig:delta_fcrpss}).

Despite the substantial improvements to SST and sea ice anomaly forecasts in both hemispheres, ocean-atmosphere coupling has a limited impact on weekly mean extratropical atmospheric variability in these experiments (figure \ref{fig:delta_fcrpss}). Two-metre temperature (T$_{2m}$) is a notable exception and is significantly improved at lead times of two to six weeks, which likely reflects a direct response to the improved SST anomalies in ENS-C compared to ENS-U for the same lead times. In contrast, week one extratropical T$_{2m}$ anomaly forecasts are slightly degraded in ENS-C relative to ENS-U, which is a consequence of increased week one biases and reduced anomaly skill concentrated over the Arctic sea ice. In coupled configurations, thermodynamic flux calculations take into account sea ice and snow thickness information from the dynamic SI$^3$ model, whereas both \uncoupled{} and the ERA5 reference assume a constant and uniform ice thickness of 1.5 m. This results in systematically cooler T$_{2m}$ over the Arctic in \coupled{} compared to both \uncoupled{} and ERA5. Since ERA5 and \uncoupled{} share the same systematic errors in T$_{2m}$ over sea ice \citep[e.g.][]{zampieri2023machine}, the apparent degradation of week one T$_{2m}$ in \coupled{} is partly an artefact of verification against ERA5. 

Other extratropical atmospheric anomaly fields that exhibit notable improvements in ENS-C relative to ENS-U include streamfunction and velocity potential at 200 hPa (figure \ref{fig:delta_fcrpss}). However, these improvements are not accompanied by robust improvements in fields that are more representative of the local extratropical circulation, such as 200 hPa zonal and meridional winds, 500 hPa geopotential height, or mean sea level pressure (figure \ref{fig:delta_fcrpss}). One important difference is that streamfunction and velocity potential are smooth, non-local fields derived from the Laplacian inversion of vorticity and divergence, respectively, which means their skill in the extratropics is also influenced by changes to divergent and rotational winds in the tropics. We therefore interpret the extratropical improvements in these fields as primarily reflecting the substantial improvements in tropical and subtropical skill, rather than an improved representation of the local extratropical circulation. 

We also see some evidence for improvements to extratropical skill for a range of tropospheric parameters at week four in the northern hemisphere (figure \ref{fig:delta_fcrpss}). However, given that comparisons for different parameters are not independent and we do not see similar improvements in weeks three or five, these apparent improvements may be a consequence of sampling uncertainty. Further investigation with a larger sample of start dates may be required for a more robust evaluation of the extratropical atmospheric response to ocean-atmosphere coupling at subseasonal timescales.

\begin{figure}[!htbp]
    \includegraphics[width=14cm]{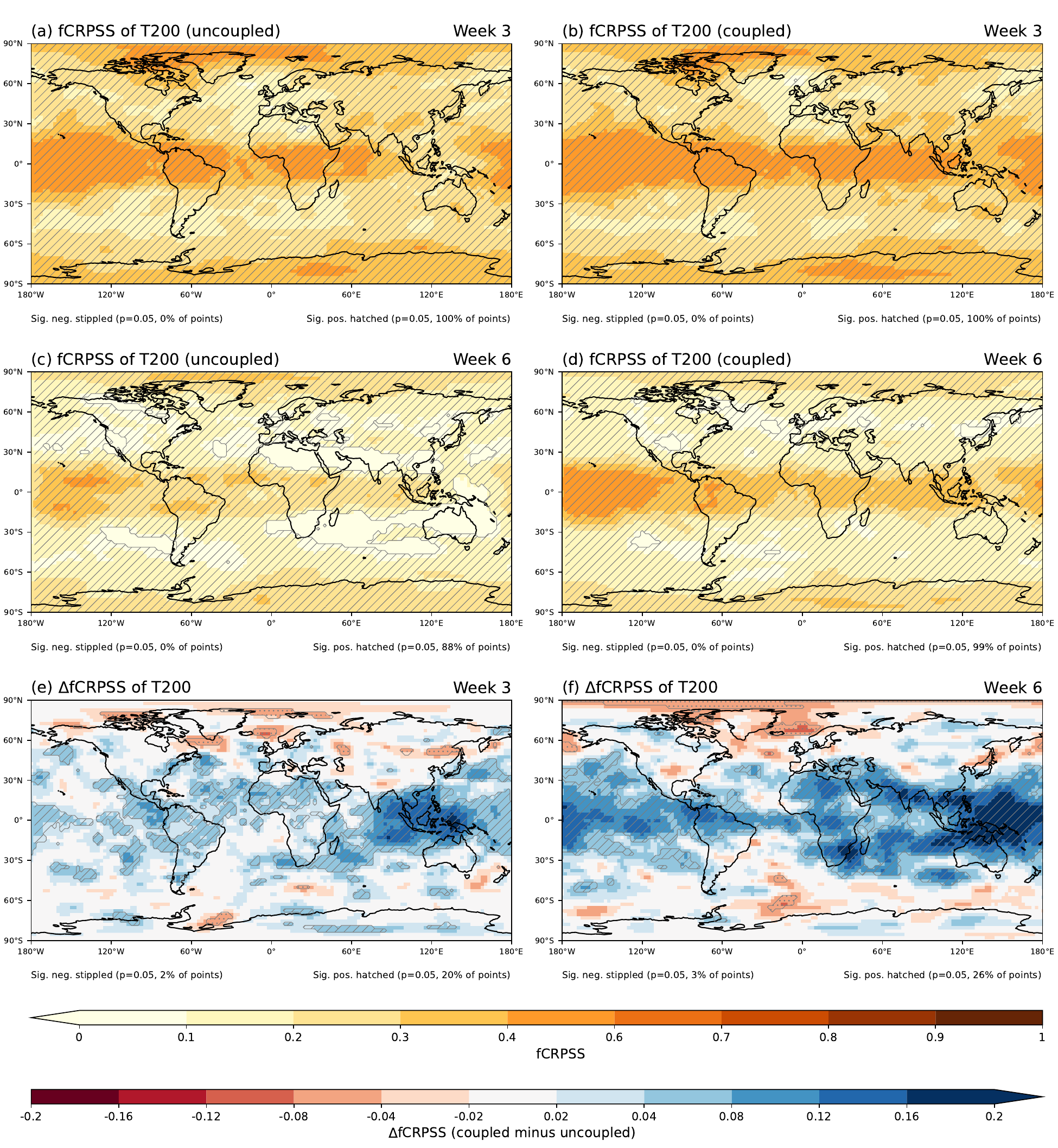}
    \centering
    \caption{(a, b) fCRPSS of week three temperature anomalies at 200 hPa in uncoupled (ENS-U) and coupled (ENS-C) subseasonal reforecasts. (c, d) As above, but for week six. (e, f) Differences in fCRPSS for weekly mean temperature anomalies at 200 hPa expressed as ENS-C minus ENS-U, such that positive values (blue shading) correspond to improvements associated with ocean-atmosphere coupling. Hatching/stippling indicates regions where estimated values are different from zero and robust to our estimates of sampling uncertainty such that the 2.5th and 97.5th percentiles of empirical bootstrap distributions have the same sign.}
    \label{fig:coupling_impact_t200}
\end{figure}

To provide a more regional evaluation of the impacts of ocean-atmosphere coupling, figure \ref{fig:coupling_impact_t200} shows changes in fCRPSS for 200 hPa temperature anomalies (T200) at lead times of three and six weeks. For a lead time of three weeks, regional T200 skill is similar in ENS-U and ENS-C (figure \ref{fig:coupling_impact_t200}a,b). In both reforecast configurations, week three skill is generally higher in the tropics than the extratropics and significantly better than climatology for 100 \% of plotted locations. Nevertheless, ocean-atmosphere coupling already has a statistically robust and positive impact on week three T200 forecasts at 20\% of locations, with the largest impacts over the maritime continent and the tropical Indian and western Pacific Oceans (figure \ref{fig:coupling_impact_t200}e). The impact of ocean-atmosphere coupling on T200 anomalies is even clearer at week six (figure \ref{fig:coupling_impact_t200}c,d). such that 99 \% of locations exceed climatological forecast skill in ENS-C compared to only 88\% of locations in ENS-U. The positive impacts of ocean-atmosphere coupling are generally restricted to within $\pm$45$^{\circ}$ latitude of the equator, with the largest impacts above the tropical warm pool in the western Pacific Ocean (figure \ref{fig:coupling_impact_t200}f). The main features of these results are consistent when the analysis is restricted to extended boreal winter (November--April) or extended boreal summer (May--October) start dates (not shown).

\begin{figure}[!htbp]
    \includegraphics[width=14cm]{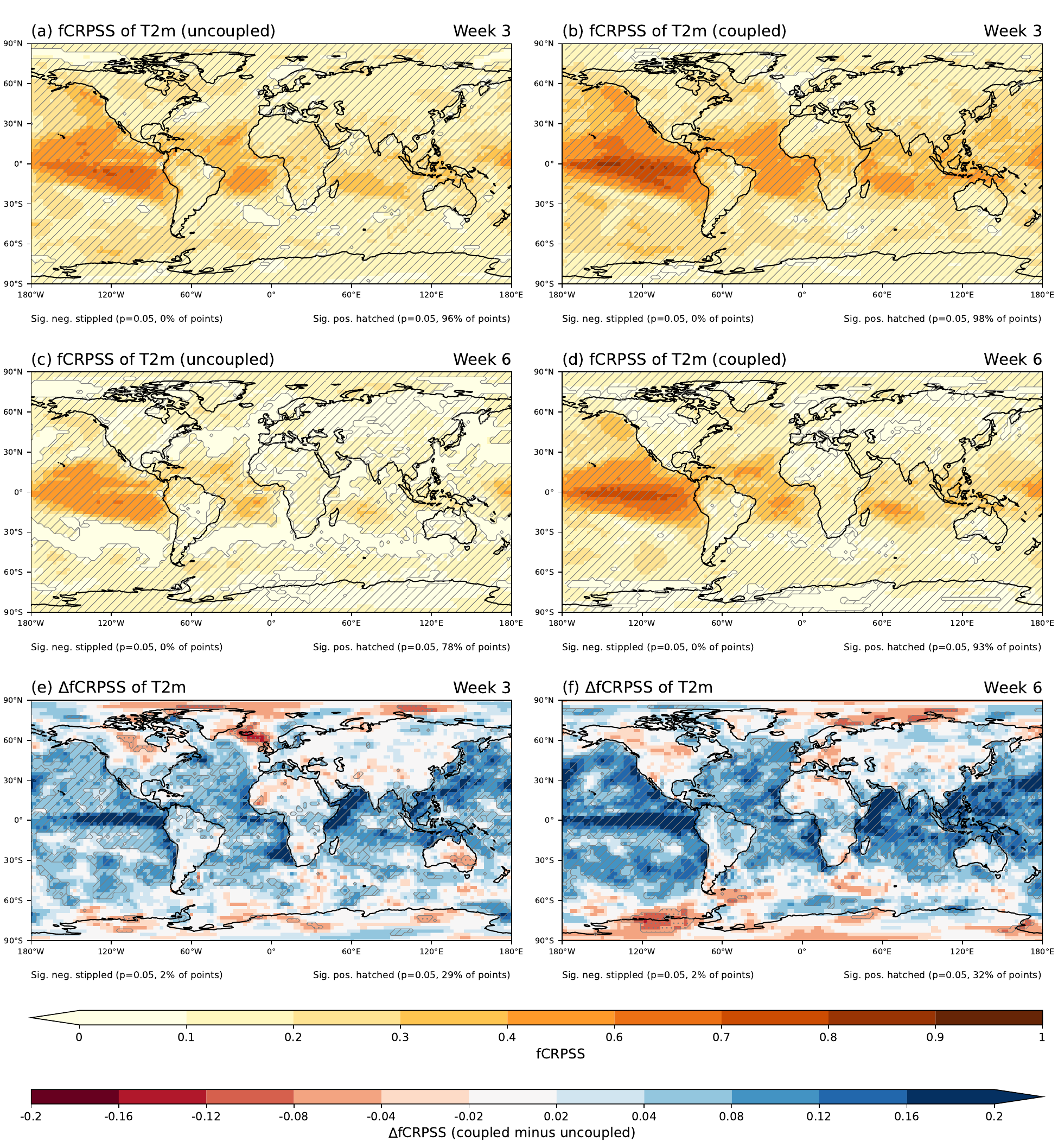}
    \centering
    \caption{As figure \ref{fig:coupling_impact_t200}, but for two-metre temperature (T$_{2m}$) anomalies.}
    \label{fig:coupling_impact_t2m}
\end{figure}

Figure \ref{fig:coupling_impact_t2m} shows the equivalent analysis for T$_{2m}$. The regional patterns of T$_{2m}$ skill are similar in ENS-U and ENS-C at both lead times, with the highest skill over the tropical oceans, and particularly the tropical eastern Pacific Ocean, where the pattern of T$_{2m}$ skill follows that of SST (figure \ref{fig:coupling_impact_t2m}a-d). However, fewer locations are significantly better than climatology in ENS-U than in ENS-C at both week three (96\% compared to 98\%) and week six (78\% compared to 93\%). The improvements in T$_{2m}$ skill associated with ocean-atmosphere coupling are found mainly over the oceans and tropical land areas, and are statistically robust at 29\% and 32\% of locations at weeks three and six, respectively (figure \ref{fig:coupling_impact_t2m}e,f). There are also some locations at higher latitudes where ocean-atmosphere coupling slightly degrades T$_{2m}$ forecasts, such as near Iceland at week three, but these account for only 2\% of locations at both lead times. As discussed above for week one, some of these apparent degradations may reflect the shared systematic errors in T$_{2m}$ over sea ice in ERA5 and \uncoupled{}.

\subsection{Impacts on the Madden-Julian Oscillation (MJO)}
\begin{figure}[!htbp]
    \includegraphics[width=14cm]{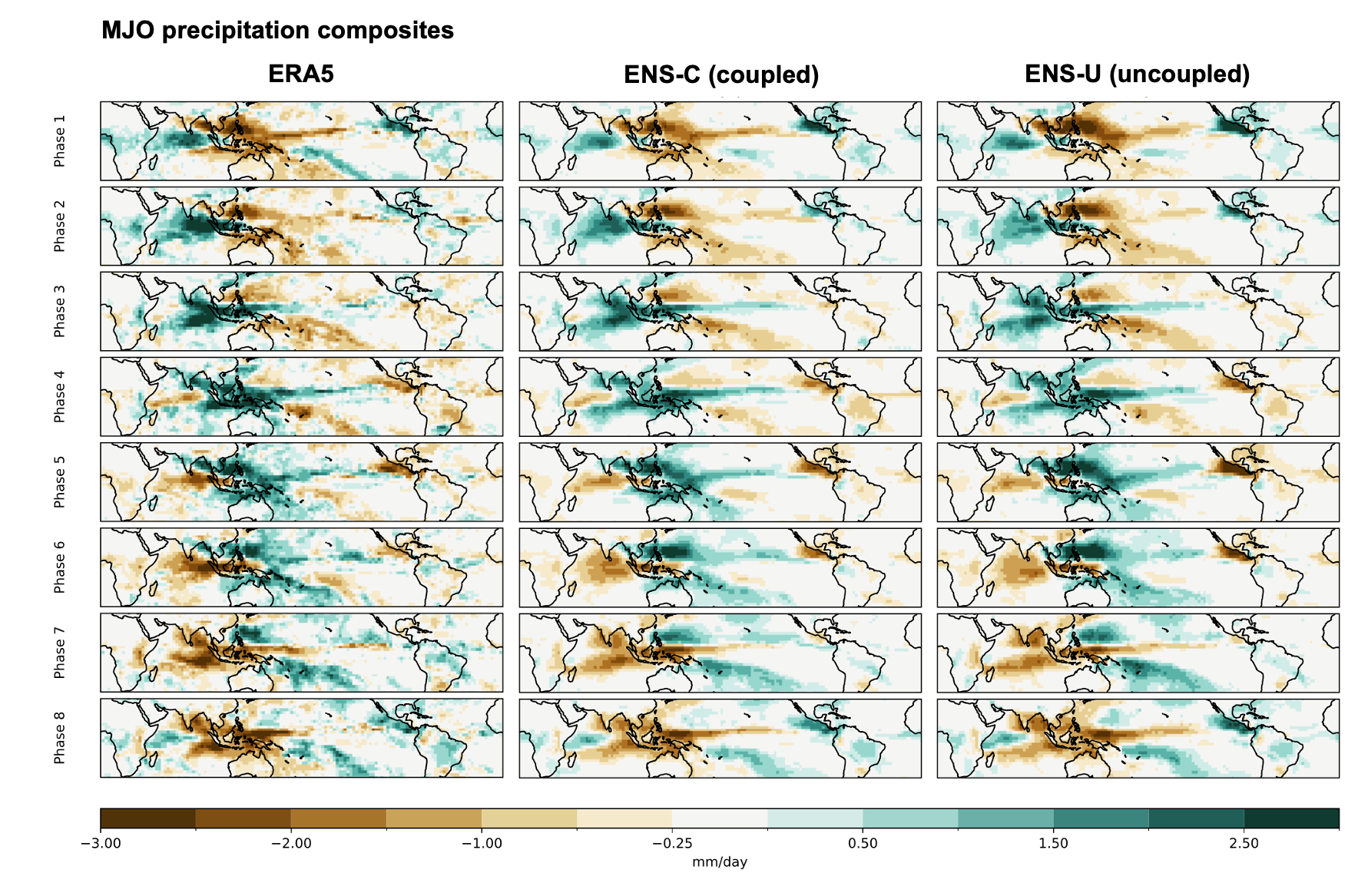}
    \centering
    \caption{Composite means of precipitation anomalies for each MJO phase for the period 2006-2023. Contributing data are selected using the MJO phase calculated separately in each forecast member and MJO events with amplitude less than one are excluded from the composite calculation. All forecast lead times are considered together and ERA5 data are subsampled to exactly match the available forecast data.}
    \label{fig:mjo_composites}
\end{figure}

\begin{figure}[!htbp]
    \includegraphics[width=14cm]{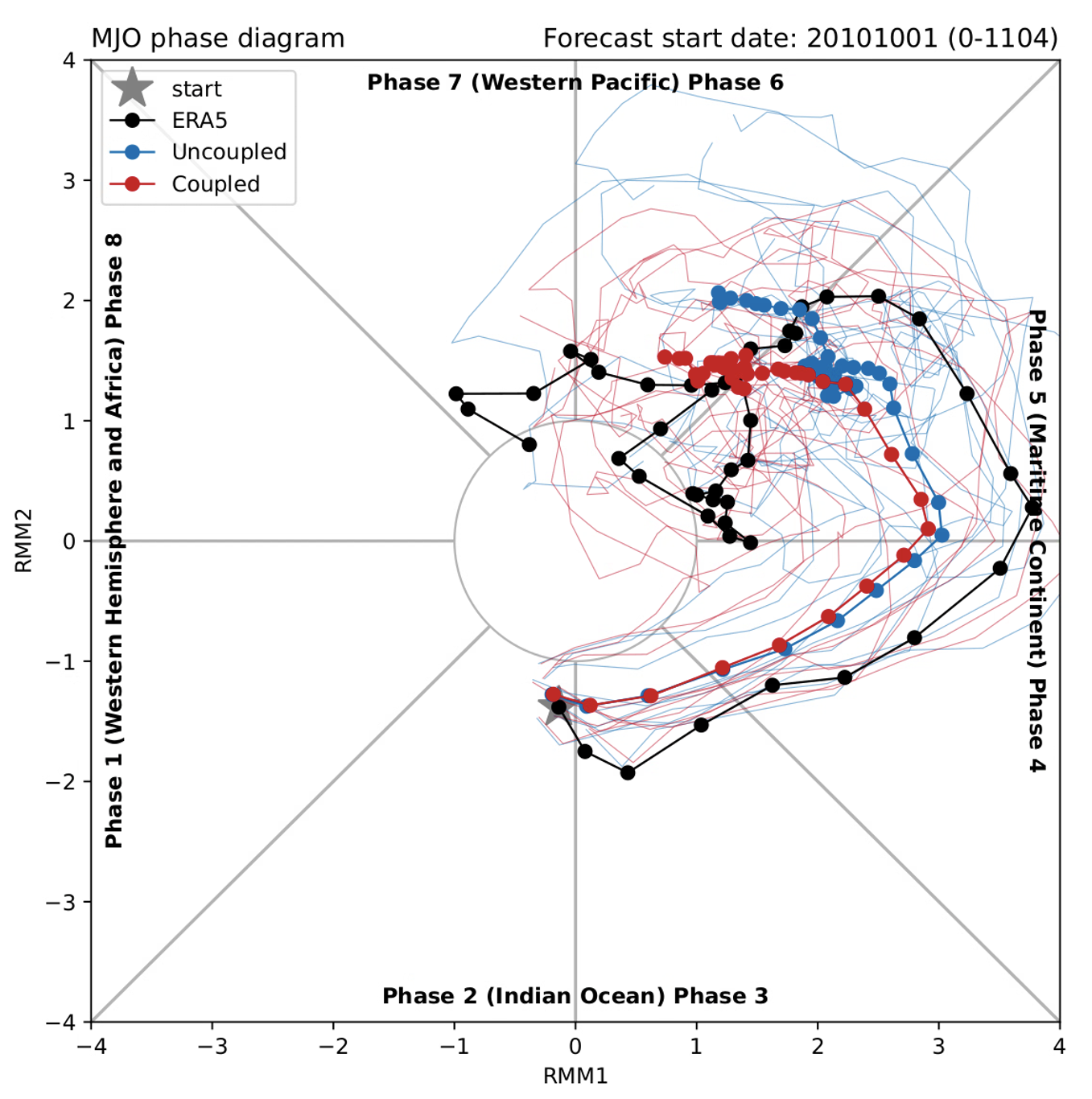}
    \centering
    \caption{A phase diagram showing the evolution of the MJO in ERA5 (black), ENS-C (red), and ENS-U (blue) for ensemble forecasts initialised on October 1st 2010. Ensemble mean and ERA5 trajectories are shown as thicker lines with solid circles marking each day of the forecast. Individual ensemble members are shown as thinner lines.}
    \label{fig:mjo_phase_diag}
\end{figure}

This section evaluates the impact of ocean-atmosphere coupling on the Madden-Julian Oscillation (MJO), which is the leading mode of intraseasonal variability in the tropics and an important source of predictability at subseasonal lead times \citep{madden1971detection, vitart2017madden}. The MJO is characterised by eastward propagating precipitation and circulation anomalies, which evolve with a characteristic timescale of 30-60 days. Figure \ref{fig:mjo_composites} shows composite means of precipitation anomalies for each MJO phase in ERA5, ENS-C, and ENS-U. In general, both reforecasts exhibit realistic MJO propagation as diagnosed from composite average precipitation anomalies such that it is difficult to identify a clear impact of ocean-atmosphere coupling. In addition, it is difficult to discriminate between coupled and uncoupled reforecast configurations when examining the evolution of specific MJO events in RMM phase space (figure \ref{fig:mjo_phase_diag}). For example, the ensemble mean and member RMM trajectories from ENS-C and ENS-U forecasts initialised on October 1st 2010 are extremely similar (figure \ref{fig:mjo_phase_diag}). In this specific example, the RMM amplitudes are slightly higher and closer to ERA5 estimates in the uncoupled ENS-U configuration. 

\begin{figure}[!htbp]
    \includegraphics[width=14cm]{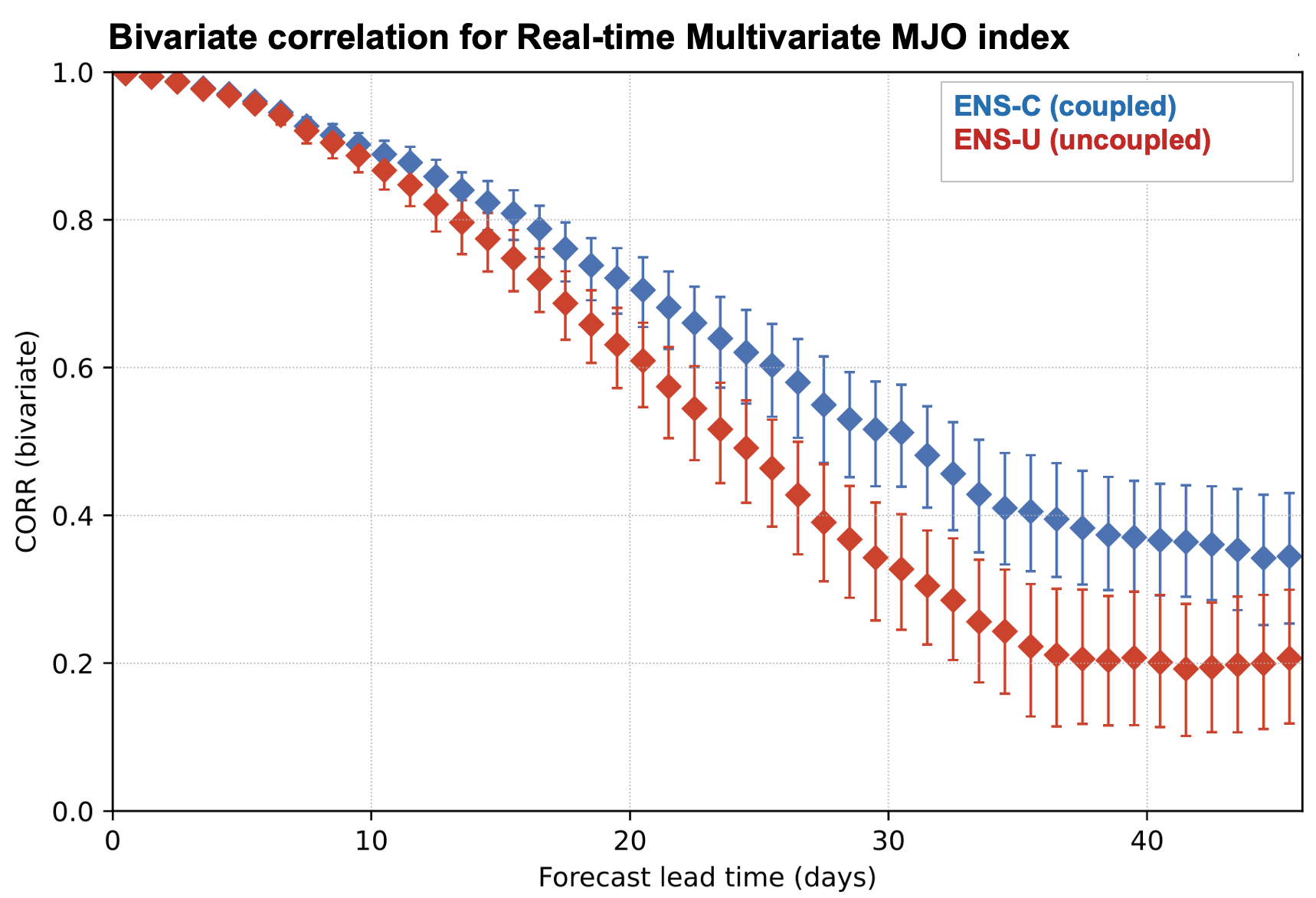}
    \centering
    \caption{Bivariate correlations for the RMM index in ENS-C (blue) and ENS-U (red) verified against indices calculated using ERA5. Error bars represent the 2.5 and 97.5th percentiles of an empirical distribution created by bootstrap resampling of the available start dates.}
    \label{fig:mjo_corrs}
\end{figure}

Nevertheless, when MJO forecast skill is assessed over all cases, we find a strong positive impact of ocean-atmosphere coupling that emerges at lead times longer than seven days (figure \ref{fig:mjo_corrs}). In particular, the lead time at which ensemble mean correlations reach a threshold value of 0.6 increases from 21 days in ENS-U to 26 days in ENS-C. This positive impact is also evident in probabilistic metrics of MJO forecast skill (not shown), such that MJO predictions from ENS-C are significantly better than ENS-U at all lead times beyond day seven. In these simulations, the improvements in MJO skill associated with the activation of ocean-atmosphere coupling correspond to faster propagation and a reduced phase bias. These positive impacts of ocean-atmosphere coupling on RMM predictability are more pronounced during the extended boreal winter (November--April), but also evident during boreal summer (May--October) start dates. Overall, these results are consistent with previous work and the interpretation of the MJO as a predominantly atmospheric mode, but with an important role for air-sea interactions and coupling between the ocean mixed layer, atmospheric convection, and moisture advection \citep{woolnough2007role, kim2010ocean, klingaman2014role, demott2015atmosphere}.

\subsection{Impacts on extratropical modes of atmospheric variability}
\begin{figure}[!htbp]
    \includegraphics[width=7cm]{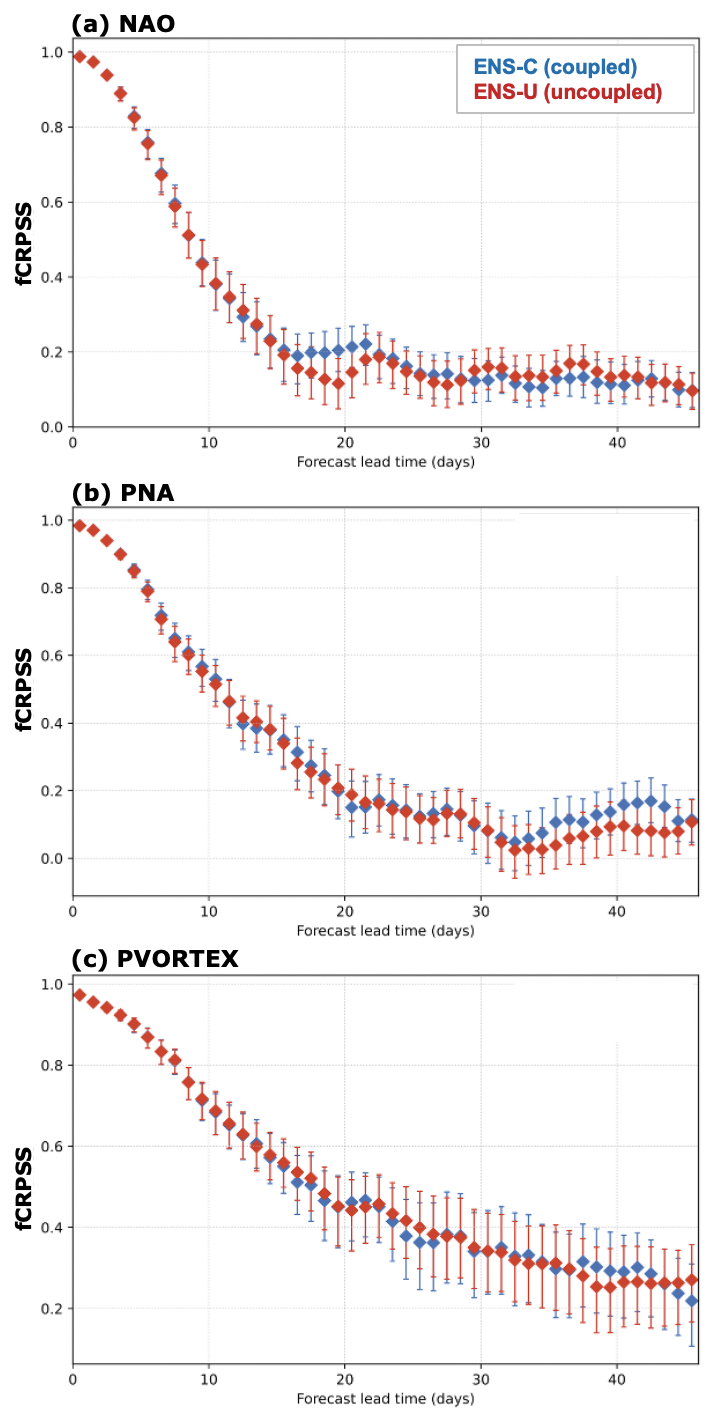}
    \centering
    \caption{Fair CRPSS of the daily mean extratropical atmospheric circulation indices described in section \ref{section:methods} in ENS-U (red) and ENS-C (blue). Error bars represent the 2.5 and 97.5th percentiles of an empirical distribution created by bootstrap resampling of the available start dates.}
    \label{fig:nao_etc}
\end{figure}

Figure \ref{fig:nao_etc} shows the impact of ocean-atmosphere coupling on three indices of large-scale atmospheric circulation in the extratropical northern hemisphere. Despite the dramatic improvements to SST and sea ice predictions in the extratropics, the subseasonal forecast skill for NAO, PNA, and PVORTEX indices is extremely similar in ENS-C and ENS-U reforecast configurations. More specifically, ENS-C and ENS-U estimates of fCRPSS are consistent within our estimated 95\% confidence intervals for almost all lead times for all three circulation indices. In addition, both coupled and uncoupled reforecast configurations provide a consistent picture of the relative predictability of these indices, such that PVORTEX $>$ PNA $>$ NAO. These results are qualitatively unchanged for other metrics of forecast skill, such as correlations (not shown), and are not sensitive to whether the analysis is restricted to extended boreal winter (November--April) or extended boreal summer (May--October) start dates. This analysis of daily mean indices is consistent with the limited impact of ocean-atmosphere coupling on extratropical weekly mean anomalies shown in figure \ref{fig:delta_fcrpss}.

\section{Impacts of an eddy-rich ocean}
\label{section:results_resolution}
\subsection{Eddy-rich initial conditions}
\begin{figure}[!htbp]
    \includegraphics[width=14cm]{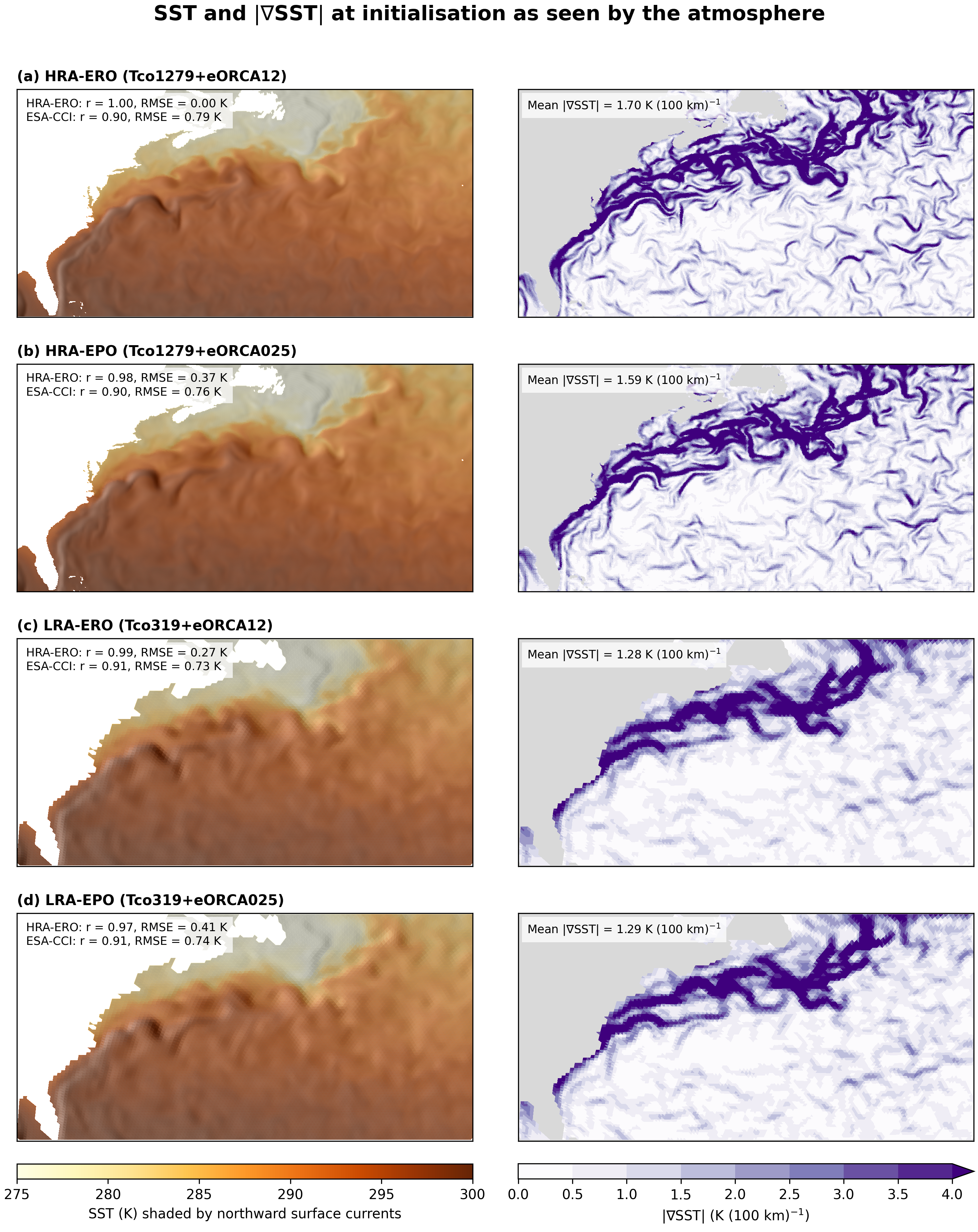}
    \centering
    \caption{SST initial conditions as seen by the IFS atmosphere in the Gulf Stream region (25$^{\circ}$N-50$^{\circ}$N, 85$^{\circ}$W-30$^{\circ}$W) for January 1st 2016 in the eddy-permitting and eddy-rich reforecast configurations described in table \ref{tab:experiments}. Left-hand panels show SST values plotted as a Delaunay triangulation of the native IFS grid such that colours represent the mean of the three IFS grid points that define the triangle, shaded using meridional surface current velocity to give a three dimensional effect and highlight Gulf Stream meanders and eddies. The annotations on the left-hand panels represent cosine-latitude weighted spatial correlations and root-mean-square differences between high-passed SSTs in each configuration, \hraero{}, and ESA-CCI observations over the plotted domain as described in section \ref{section:results_resolution}. Right-hand panels show the magnitude of the horizontal SST gradient, calculated on the native IFS grid using a finite-element gradient operator. Land points and coastal grid points whose gradients depend on land values are excluded (grey shading). The annotations on the right-hand panels represent area-weighted mean SST gradient magnitudes over the plotted domain.}
    \label{fig:gulf_stream}
\end{figure}

In this section, we combine results from the eddy-permitting (\lraepo{}, \hraepo{}) and eddy-rich (\lraero{}, \hraero{}) reforecast configurations described in table \ref{tab:experiments} to evaluate the impact of increasing ocean resolution on subseasonal predictability. To begin, we confirm that our ocean initialisation procedure (see Appendix A) is behaving as expected such that ocean mesoscale features are represented coherently across eddy-rich and eddy-permitting reforecast initial conditions (figure \ref{fig:gulf_stream}). Despite some differences in the shape and intensity of individual eddies, all reforecast configurations capture the main features of the Gulf Stream SST front and its extension at initialisation time, including the location of meanders and sharp horizontal gradients.

To quantify this agreement against satellite observations, we compare the initial SSTs seen by the atmosphere with the ESA-CCI/C3S SST analysis for the same date. All fields are interpolated to a common 0.1$^{\circ}\times$0.1$^{\circ}$ latitude-longitude grid and high-pass filtered with the 6$^{\circ}$ latitude by 18$^{\circ}$ longitude boxcar filter described in section \ref{section:methods_scale}, such that the comparison focuses on mesoscale features rather than the large-scale meridional SST gradient. Over the domain shown in figure \ref{fig:gulf_stream}, cosine-latitude weighted spatial correlations between high-passed initial SSTs and ESA-CCI are 0.90 to 0.91 in all four configurations, with root-mean-square differences (RMSDs) of 0.73 to 0.79 K. High-passed initial SSTs are also highly consistent across configurations, with correlations of 0.97 to 0.99 and RMSD of 0.27 to 0.41 K relative to \hraero{}. All correlations for the plotted start date are in the range 0.99 to 1.00 when calculated from raw SSTs without high-pass filtering. 

Mesoscale SST anomalies at initialisation time are therefore well constrained by observations in both eddy-permitting and eddy-rich configurations, regardless of atmospheric resolution. Nevertheless, the observations available to constrain individual ocean eddies, and particularly their subsurface structure, are sparse in time and space such that different ocean analyses can place individual eddies in different locations \citep{jacobs2023adapting}. This uncertainty would ideally be sampled using an ensemble of eddy-rich ocean state estimates, analogous to the eddy-permitting ORAS6 ensemble reanalysis used to initialise \coupled{} \citep{zuo2024ecmwf}. However, our evaluation of eddy-rich configurations is based on single-member deterministic reforecasts and thus uncertainty in the location and vertical structure of individual eddies, and its potential impact on subseasonal forecasts, is not assessed in this study.

The comparisons in figure \ref{fig:gulf_stream} also highlight that the sharpness of the SST fronts and gradients experienced by the atmosphere can be more sensitive to the horizontal resolution of the atmosphere than the ocean (right-hand panels of figure \ref{fig:gulf_stream}). Area-weighted mean SST gradient magnitudes over the plotted domain are 1.70 and 1.59 K (100 km)$^{-1}$ in \hraero{} and \hraepo{}, respectively, but only 1.28 and 1.29 K (100 km)$^{-1}$ in \lraero{} and \lraepo{}. With the Tco1279 atmosphere, the eddy-rich ocean provides more numerous and sharper SST filaments than the eddy-permitting ocean (figure \ref{fig:gulf_stream}a vs \ref{fig:gulf_stream}b). In contrast, the higher-resolution ocean state is substantially smoother when sampled by the Tco319 atmosphere, such that its SST gradients closely resemble those of the lower-resolution ocean state viewed at the same atmospheric resolution (figure \ref{fig:gulf_stream}c vs \ref{fig:gulf_stream}d).

\subsection{Impacts on the mean state}
Figures \ref{fig:eddy_impact_tco319} and \ref{fig:eddy_impact_tco1279} summarise the impact of increasing ocean resolution from the eddy-permitting to eddy-rich regime in single-member subseasonal reforecasts for different variables, lead times, and regions. The impact on the climatological mean state is consistent across both atmospheric resolutions, with small but statistically robust improvements to SST, sea ice cover, and dynamic sea level in the extratropical northern hemisphere and small improvements to mean sea level in the tropics. Otherwise, we find very limited evidence for statistically robust changes in the atmospheric mean state in response to increased ocean resolution in these single-member experiments (figures \ref{fig:eddy_impact_tco319} and \ref{fig:eddy_impact_tco1279}). We find similar results for the extratropical atmosphere in the southern hemisphere, albeit with some evidence for degradations to climatological T$_{2m}$ associated with increased sea ice cover biases in our eddy-rich ocean configurations (not shown). 

\begin{figure}[!htbp]
    \includegraphics[width=14cm]{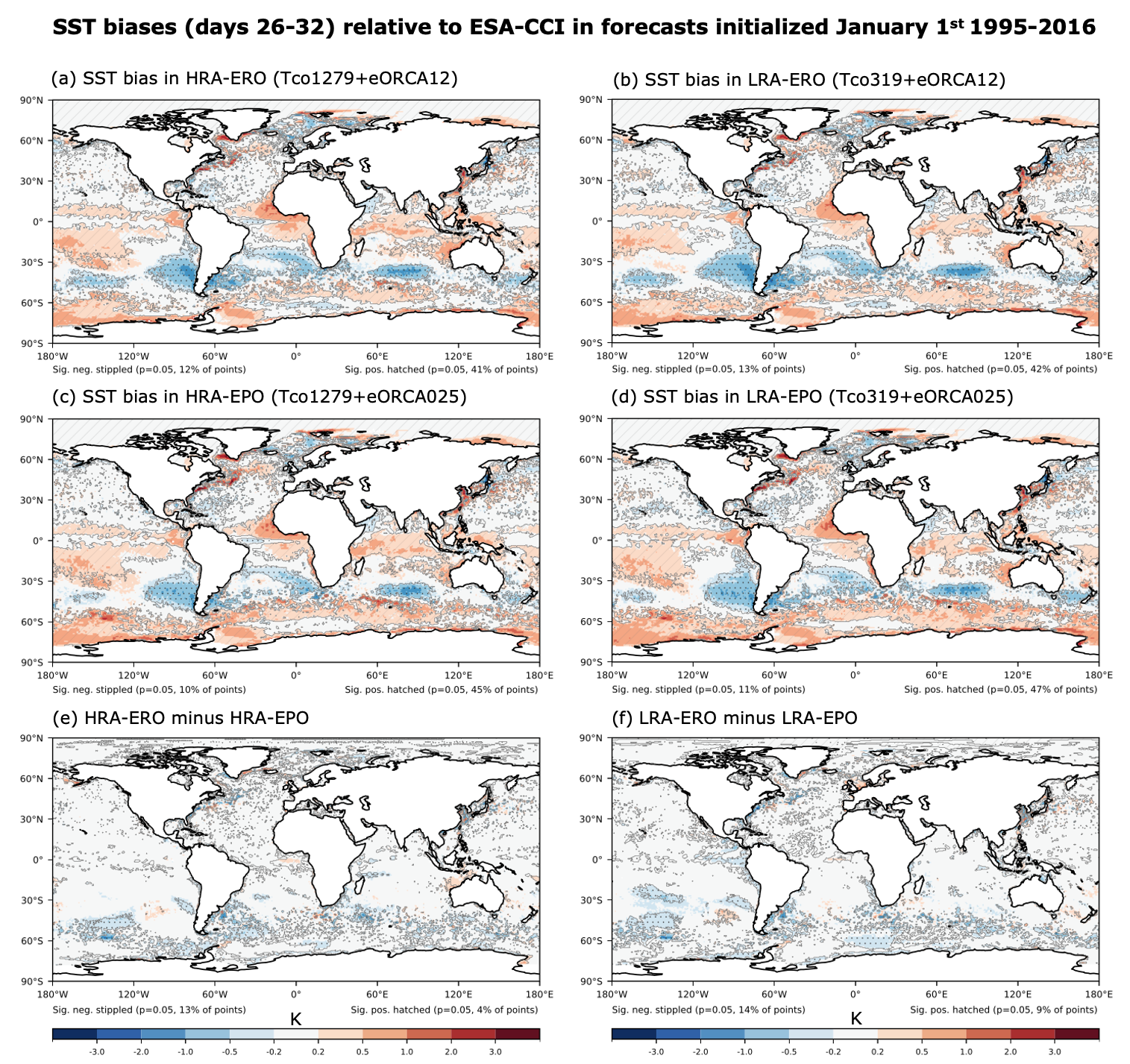}
    \centering
    \caption{(a-d) Week 4$\frac{1}{2}$ (i.e. days 26-32) SST biases in eddy-permitting and eddy-rich reforecasts initialised on January 1st 1995-2016 relative to ESA-CCI SST \citep{merchant2019satellite}. (e,f) Difference between January 1st SST climatologies in eddy-rich and eddy-permitting reforecast configurations. Hatching/stippling indicates regions where estimated values are different from zero and robust to our estimates of sampling uncertainty such that the 2.5th and 97.5th percentiles of empirical bootstrap distributions have the same sign.}
    \label{fig:eddy_impact_sst_biases}
\end{figure}

The regional details and large-scale structure of SST biases are extremely similar for all four combinations of ocean and atmosphere resolution (figure \ref{fig:eddy_impact_sst_biases}a-d). For example, week 4$\frac{1}{2}$ forecasts (i.e. days 26 to 32) initialised on January 1st all exhibit comparable positive SST biases associated with the Gulf Stream, East Greenland Current, tropical oceans, and the Southern Ocean south of 50 $^{\circ}$S. There is also strong agreement on the location of negative SST biases, particularly in the mid-latitudes oceans of the Southern Hemisphere. Nevertheless, we find some evidence of improvements to regional SST biases in our eddy-rich reforecast configurations, including a widespread cooling of 0.2-0.5 K in the Southern Ocean and a slight reduction to the magnitude of SST biases in the Gulf Stream region for January 1st start dates (figure \ref{fig:eddy_impact_sst_biases}e,f). These differences are statistically robust and consistent across both atmospheric resolutions, but are modest compared to the overall bias magnitudes, which remain comparable across all configurations. 

\subsection{Impacts on weekly mean forecast anomalies}
\label{section:resolution_anomaly_impact}
\begin{figure}[!htbp]
    \includegraphics[width=14cm]{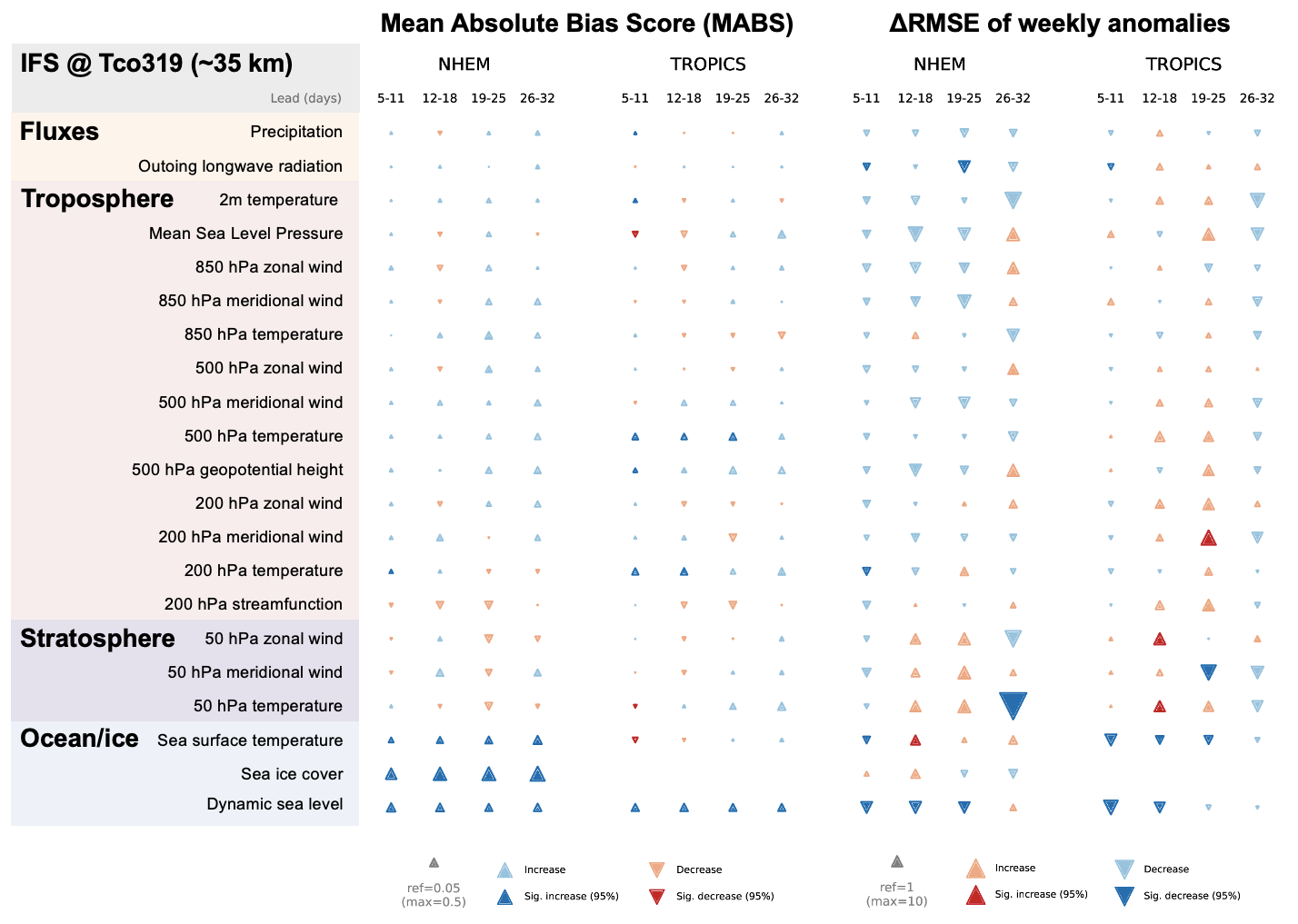}
    \centering
    \caption{Score cards summarising the subseasonal impact of increasing ocean horizontal resolution from $\sim$25 km to $\sim$8 km while coupled to the IFS Tco319 atmosphere, where blue triangles indicate increased MABS or reduced RMSE in \lraero{} relative to \lraepo{}. RMSE differences are scaled such that $\Delta \mathrm{RMSE} = \frac{\mathrm{RMSE}_{\mathrm{ERO}}- \mathrm{RMSE}_{\mathrm{EPO}}}{\mathrm{RMSE}_{\mathrm{EPO}}}$. Score differences are calculated for the following regions: NHEM (30$^{\circ}$N-90$^{\circ}$N) and TROPICS (30$^{\circ}$S-30$^{\circ}$N). Symbol areas are proportional to the magnitude of changes and absolute values can be inferred by comparison with the area of the grey reference triangles. Darker colours indicate that bootstrap-derived 95\% confidence intervals do not intersect zero.}
    \label{fig:eddy_impact_tco319}
\end{figure}

\begin{figure}[!htbp]
    \includegraphics[width=14cm]{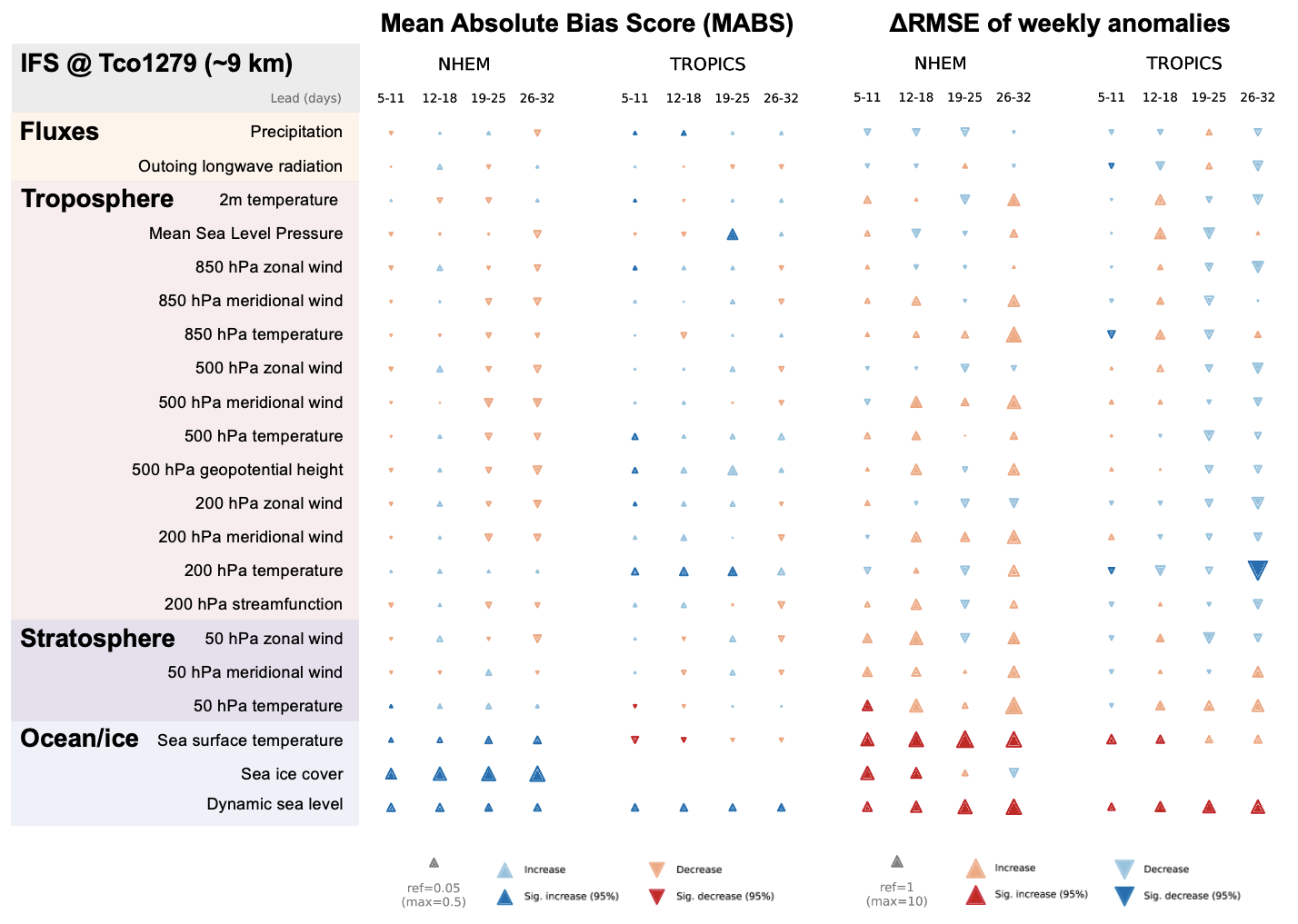}
    \centering
    \caption{As figure \ref{fig:eddy_impact_tco319}, but for reforecasts using the IFS Tco1279 atmosphere. Blue triangles indicate increased MABS or reduced RMSE in \hraero{} relative to \hraepo{}.}
    \label{fig:eddy_impact_tco1279}
\end{figure}

The impacts of ocean resolution on weekly mean anomaly forecast skill diagnosed using differences in anomaly-based RMSE are also summarised in figures \ref{fig:eddy_impact_tco319} and \ref{fig:eddy_impact_tco1279}. In general, we find that the weekly mean RMSE of atmospheric fields is extremely similar in eddy-rich and eddy-permitting reforecast configurations. It is possible to identify some combinations of region, lead time, and atmospheric resolution where increased ocean resolution is associated with seemingly significant reductions in RMSE (e.g. 50 hPa temperature in the Northern Hemisphere for days 26-32 in LRA-ERO relative to LRA-EPO). However, when balancing the evidence across different atmospheric resolutions, variables, and lead times, we do not find strong evidence that these eddy-rich ocean configurations provide substantial or statistically robust improvements to atmospheric forecast skill at subseasonal lead times (figures \ref{fig:eddy_impact_tco319} and \ref{fig:eddy_impact_tco1279}). Nevertheless, it remains possible that modest improvements, not detectable in these single-member experiments, could emerge with ensemble reforecasts and/or a larger sample of start dates.

The Years of the Maritime Continent initiative \citep{yoneyama2020years} emphasised that air-sea interactions and associated atmospheric variability in this region may be particularly sensitive to model resolution due to the presence of complex coastlines, shallow seas, and narrow straits. However, we find no consistent impact of increased ocean model resolution on deterministic MJO forecast skill in these single member reforecasts (not shown). Bivariate correlations of the RMM index are almost identical in all four configurations for the first three weeks, and fall below 0.6 after 22 to 24 days. At longer lead times, eddy-rich reforecasts have slightly higher correlations than eddy-permitting reforecasts when coupled to the Tco1279 atmosphere (by up to 0.05), but slightly lower correlations when coupled to the Tco319 atmosphere (by up to 0.03). However, bootstrap-derived 95\% confidence intervals for these differences include zero for all but one subseasonal lead time (day 23 when comparing HRA-ERO and HRA-EPO).

In contrast, the impact of increased ocean resolution on the RMSE of weekly mean ocean and sea ice anomalies seems to be sensitive to the horizontal resolution of the atmosphere model (figures \ref{fig:eddy_impact_tco319} and \ref{fig:eddy_impact_tco1279}). For example, when coupled to the lower-resolution Tco319 atmosphere, the eddy-rich ocean configuration has an either slightly positive or neutral impact on weekly mean SST, sea ice cover, and sea level anomalies in the tropics and extratropical Northern Hemisphere (figure \ref{fig:eddy_impact_tco319}). In contrast, the eddy-rich ocean configuration seems to significantly increase the RMSE of the same fields when coupled to the higher-resolution Tco1279 atmosphere (figure \ref{fig:eddy_impact_tco1279}). The absolute magnitudes of the SST RMSE differences are very small, and comparable to (or less than) the median per-pixel uncertainty of 0.18 K that is estimated for the ESA-CCI SST product \citep{merchant2019satellite}. Nevertheless, given these signals are the main detectable impact of the increased ocean resolution, we will try to understand their origins. 

There are two possible explanations for this sensitivity to atmospheric resolution: (1) A physical interpretation, in which the relative performance of eddy-rich and eddy-permitting NEMO4-SI$^3$ configurations is sensitive to the details of the atmospheric representation. (2) A statistical interpretation, in which our assessment of RMSE changes in single-member deterministic forecasts is compromised by the so-called double penalty effect \citep{lledo2023scale}. For deterministic metrics such as RMSE, forecasts that predict small-scale features with sharp gradients (e.g. convective precipitation cells or ocean eddies) are penalised twice when the simulated feature does not match the corresponding feature in the verification dataset: once for missing the feature in the correct location at the correct time, and again for having a similar feature elsewhere. In contrast, a smoother and less realistic forecast will only be penalised once for missing the observed feature. This effect is compounded by our evaluation of ocean fields archived on an intermediate IFS grid, which influences the effective resolution of the ocean features seen by the atmosphere (figure \ref{fig:gulf_stream}).

\begin{figure}[!htbp]
    \includegraphics[width=14cm]{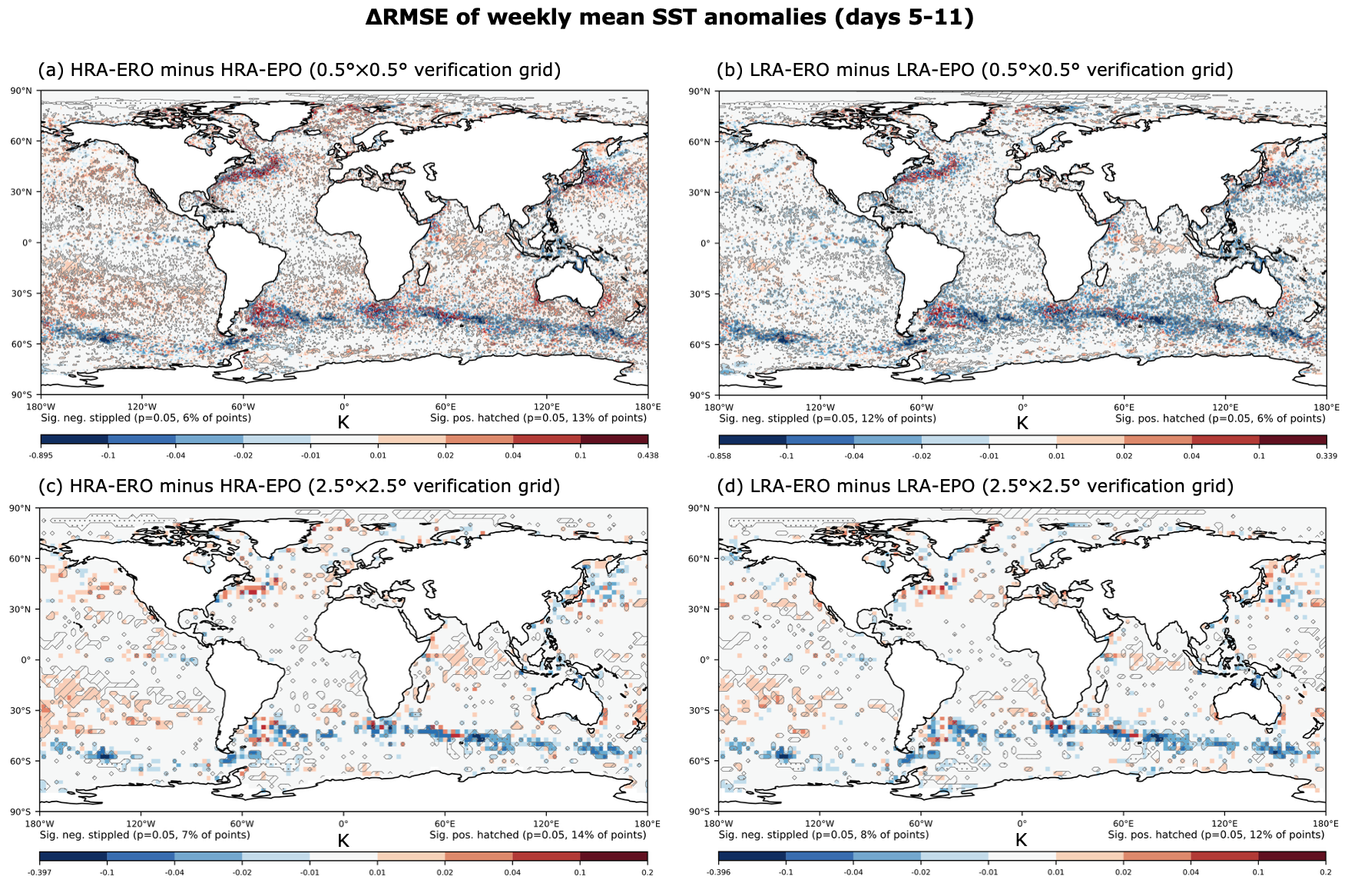}
    \centering
    \caption{(a,b) RMSE differences for week 1$\frac{1}{2}$ (i.e. days 5-11) SST anomalies estimated after conservative interpolation of all data to a common $0.5^{\circ}\times0.5^{\circ}$ latitude-longitude grid. (c,d) As above, but after conservative interpolation of all data to a common $2.5^{\circ}\times2.5^{\circ}$ latitude-longitude grid. Negative values (blue shading) correspond to reduced RMSE in eddy-rich ocean configurations. Hatching/stippling indicates regions where estimated values are different from zero and robust to our estimates of sampling uncertainty such that the 2.5th and 97.5th percentiles of empirical bootstrap distributions have the same sign.}
    \label{fig:eddy_impact_sst_rmse}
\end{figure}

A simple way to investigate the potential influence of double-penalty effects is to evaluate the scale-sensitivity of RMSE estimates using averages over different domain sizes. Figure \ref{fig:eddy_impact_sst_rmse} shows RMSE differences for week 1$\frac{1}{2}$ SST anomalies after conservative interpolation of all data to either a $0.5^{\circ}\times0.5^{\circ}$ or $2.5^{\circ}\times2.5^{\circ}$ latitude-longitude grid. The use of conservative interpolation, which can be considered a weighted average of contributing locations, means that RMSE calculated on the coarser grid is less sensitive to the exact position of individual ocean eddies and more representative of errors in the large scale ocean state. A more robust way to evaluate these small differences would be to perform ensemble simulations and apply probabilistic evaluation metrics (e.g. fCRPS) in observation space, which would simultaneously suppress the influence of double-penalty effects and limit the influence of interpolation artefacts.

When evaluated on a $0.5^{\circ}\times0.5^{\circ}$ grid, the RMSE difference maps shown in figures \ref{fig:eddy_impact_sst_rmse}a and \ref{fig:eddy_impact_sst_rmse}b are consistent with the summary score cards shown in figures \ref{fig:eddy_impact_tco319} and \ref{fig:eddy_impact_tco1279}. RMSE differences evaluated using the lower resolution atmosphere (LRA-ERO minus LRA-EPO) give a more positive impression of the impacts from an eddy-rich ocean, with 12\% of comparisons indicating a significant reduction of RMSE and 6\% of comparisons indicating a significant increase in RMSE. In contrast, RMSE differences evaluated using the higher-resolution atmosphere (HRA-ERO minus HRA-EPO) provide a more negative impression, with 6\% of comparisons indicating a significant reduction of RMSE and 13\% of comparisons indicating a significant increase in RMSE. However, RMSE differences estimated on the coarser $2.5^{\circ}\times2.5^{\circ}$ grid provide a more consistent impression of the change in RMSE associated with increased ocean resolution (7-8\% of comparisons indicating a significant reduction of RMSE and 12-14\% of comparisons indicating a significant increase in RMSE). In this example, we find that the RMSE of week 1$\frac{1}{2}$ SST anomalies is generally reduced along the Antarctic polar front and slightly increased in the Equatorial Indian Ocean, tropical Pacific Ocean, and the Gulf Stream region. Based on this analysis, the increased RMSE in the low-latitude regions seems to be a consequence of increases in the large-scale SST variance in the eddy-rich ocean configurations (not shown). Nevertheless, the resulting changes in RMSE are extremely small and may not be robust when evaluated using a different SST verification dataset.

\subsection{Scale-dependent evaluation}
\label{section:scale_dependent}
The deterministic metrics presented in \ref{section:resolution_anomaly_impact} provide an incomplete view of the impacts of increased ocean resolution as they are dominated by differences in the large-scale state and are influenced by the double-penalty effects at small scales. In this section, we complement the score-based evaluation above with the scale-dependent process diagnostics that are described in section \ref{section:methods_scale}. Firstly, we evaluate the local relationship between near-surface wind speed and SST at the ocean mesoscale, where the imprint of ocean eddies and fronts on the marine atmospheric boundary layer is expected to be strongest. Secondly, we evaluate whether increased ocean resolution modifies the distribution of tropospheric temperature variability and kinetic energy across spatial scales.

\subsubsection{Air-sea interactions at the ocean mesoscale}
\label{section:air_sea_regressions}
\begin{figure}[!htbp]
    \includegraphics[width=14cm]{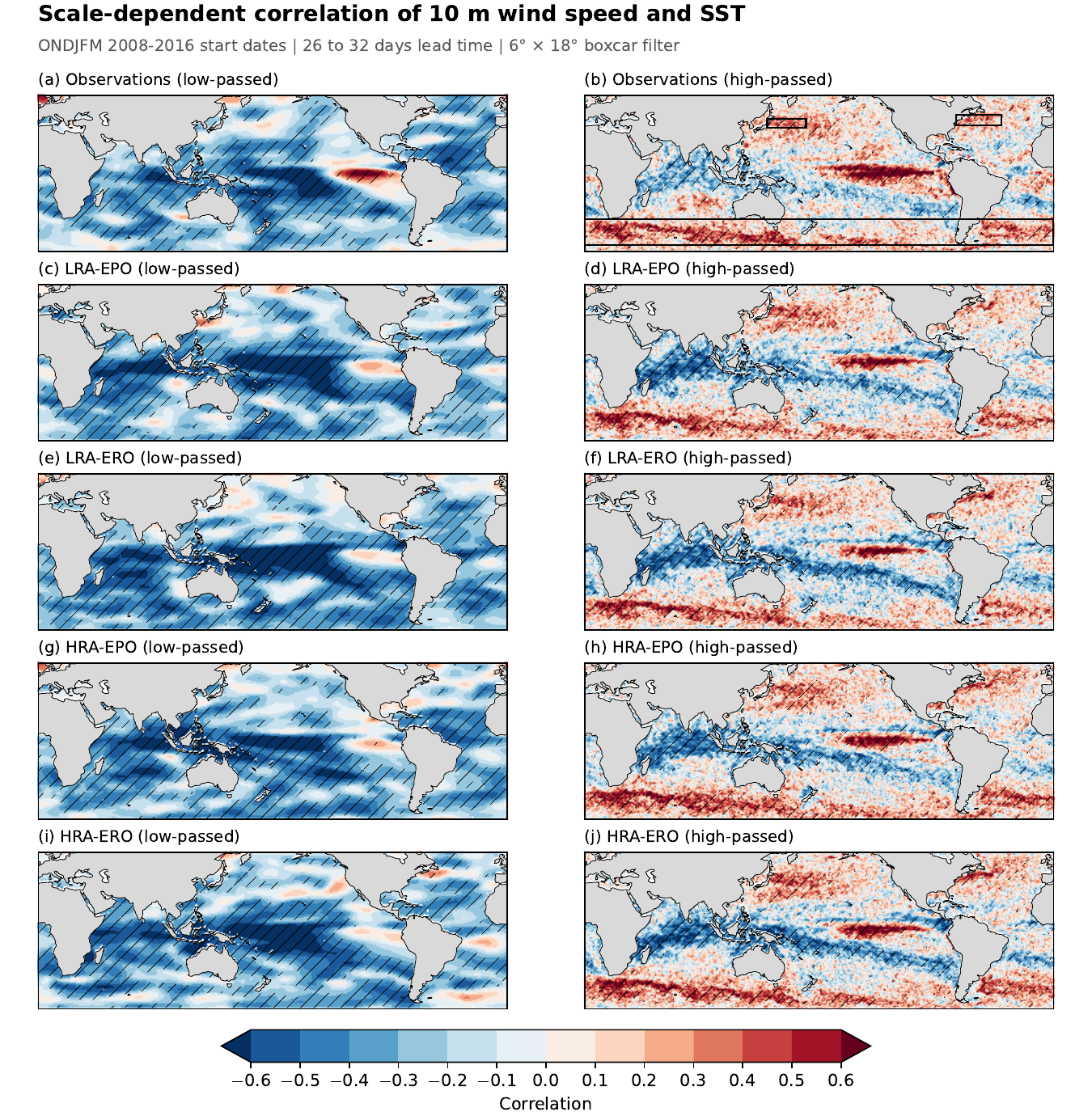}
    \centering
    \caption{Local correlations between weekly mean 10 m wind speed and SST anomalies for the period corresponding to days 26 to 32 of reforecasts initialised on the first of each month from October to March (ONDJFM) between 2008 and 2016. The left column shows correlations for low-passed anomalies obtained with a 6$^{\circ}$ latitude by 18$^{\circ}$ longitude boxcar filter and the right column shows correlations for the corresponding high-passed anomalies. (a,b) Observation-based estimates from ESA-CCI SSTs and Copernicus Marine Service scatterometer-corrected winds. (c-j) Eddy-permitting and eddy-rich reforecasts with Tco319 (\lraepo{}, \lraero{}) and Tco1279 (\hraepo{}, \hraero{}) atmospheres. Hatching indicates regions where 95\% confidence intervals estimated by bootstrap resampling over start dates do not include zero. Boxes in (b) show the regions used to calculate the regional mean regression slopes in table \ref{tab:wind_sst_slopes}.}
    \label{fig:wind_sst_correlation}
\end{figure}

Figure \ref{fig:wind_sst_correlation} shows scale-dependent local correlations between weekly mean 10 m wind speed and SST anomalies for the period corresponding to days 26 to 32 of reforecasts initialised during the extended boreal winter season (October-March). In the observation-based estimates, low-passed anomalies are negatively correlated over almost all of the ice-free ocean, with the strongest negative correlations over the tropical Indian Ocean and western Pacific (figure \ref{fig:wind_sst_correlation}a). This large-scale relationship reflects the dominant influence of the atmosphere on the ocean, in which stronger winds cool the ocean surface through enhanced turbulent heat fluxes and vertical mixing \citep{small2008air, chelton2010coupled}. In contrast, high-passed anomalies are positively correlated over the eastern equatorial Pacific and the extratropical oceans, including the Southern Ocean and the extensions of the western boundary currents (figure \ref{fig:wind_sst_correlation}b). These positive correlations are the characteristic signature of the ocean forcing the atmosphere at the ocean mesoscale, as warm SST anomalies destabilise the marine atmospheric boundary layer and enhance the downward mixing of momentum towards the surface \citep{small2008air, chelton2010coupled, bryan2010frontal}. The strongest positive correlations are found over the eastern equatorial Pacific, where tropical instability waves generate large SST anomalies along the equatorial cold tongue \citep{chelton2010coupled}. Similar patterns exist for the extended boreal summer season (April-September), including the existence of positive high-passed correlations over the eastern equatorial Pacific, the Southern Ocean, and the cores of the Gulf Stream and Kuroshio extensions (not shown).

All four reforecast configurations reproduce the main features of the observed relationships for both low-passed and high-passed anomalies (figure \ref{fig:wind_sst_correlation}c-j). The strength of the mesoscale coupling over the extratropical oceans is also broadly consistent with the observation-based estimates. For example, regional mean regression slopes of high-passed wind speed on high-passed SST anomalies during boreal winter range from 0.15 to 0.17 m s$^{-1}$ K$^{-1}$ in the reforecasts compared with 0.15 m s$^{-1}$ K$^{-1}$ in the observation-based estimate over the Gulf Stream, 0.18 to 0.20 compared with 0.20 m s$^{-1}$ K$^{-1}$ over the Kuroshio, and 0.12 to 0.18 compared with 0.18 m s$^{-1}$ K$^{-1}$ over the Southern Ocean (table \ref{tab:wind_sst_slopes}). In contrast, all reforecast configurations seem to underestimate the extent and strength of positive correlations in the eastern equatorial Pacific. During the boreal summer, regional mean slopes for high-passed anomalies remain broadly consistent with the observation-based estimates, ranging from 0.03 to 0.11 m s$^{-1}$ K$^{-1}$ compared with 0.07 m s$^{-1}$ K$^{-1}$ over the Gulf Stream, 0.10 to 0.14 compared with 0.13 m s$^{-1}$ K$^{-1}$ over the Kuroshio, and 0.26 to 0.29 compared with 0.29 m s$^{-1}$ K$^{-1}$ over the Southern Ocean (table \ref{tab:wind_sst_slopes}). 

The impact of horizontal resolution on regional mean slope coefficients is generally small, with two notable exceptions (table \ref{tab:wind_sst_slopes}). Firstly, regional mean slopes over the Southern Ocean during boreal winter increase from 0.12 m s$^{-1}$ K$^{-1}$ in both Tco319 configurations (\lraepo{} and \lraero{}) to 0.17 to 0.18 m s$^{-1}$ K$^{-1}$ in both Tco1279 configurations (\hraepo{} and \hraero{}), irrespective of ocean resolution. This sensitivity to atmospheric resolution is consistent with figure \ref{fig:gulf_stream}, which shows that the effective resolution of the SST experienced by the atmosphere is partly determined by the atmospheric grid. Secondly, regional mean slopes over the Gulf Stream during boreal summer increase from 0.03 m s$^{-1}$ K$^{-1}$ in \lraepo{} to 0.09 m s$^{-1}$ K$^{-1}$ in \lraero{}, which is the most notable impact of increased ocean resolution. However, this increase is not seen with the Tco1279 atmosphere (0.11 and 0.10 m s$^{-1}$ K$^{-1}$ for \hraepo{} and \hraero{}, respectively), such that \lraepo{} is the only configuration with weaker summertime coupling over the Gulf Stream than the observation-based estimate (0.07 m s$^{-1}$ K$^{-1}$). The relationships diagnosed for days 26 to 32 are almost identical to those for days 5 to 11 (not shown), which indicates that they are established within the first weeks of the forecast and reflect the representation of air-sea interactions in the coupled model rather than the evolution of forecast errors. 

\subsubsection{Atmospheric variance and kinetic energy spectra}
\label{section:spectra}
\begin{figure}[!htbp]
    \includegraphics[width=14cm]{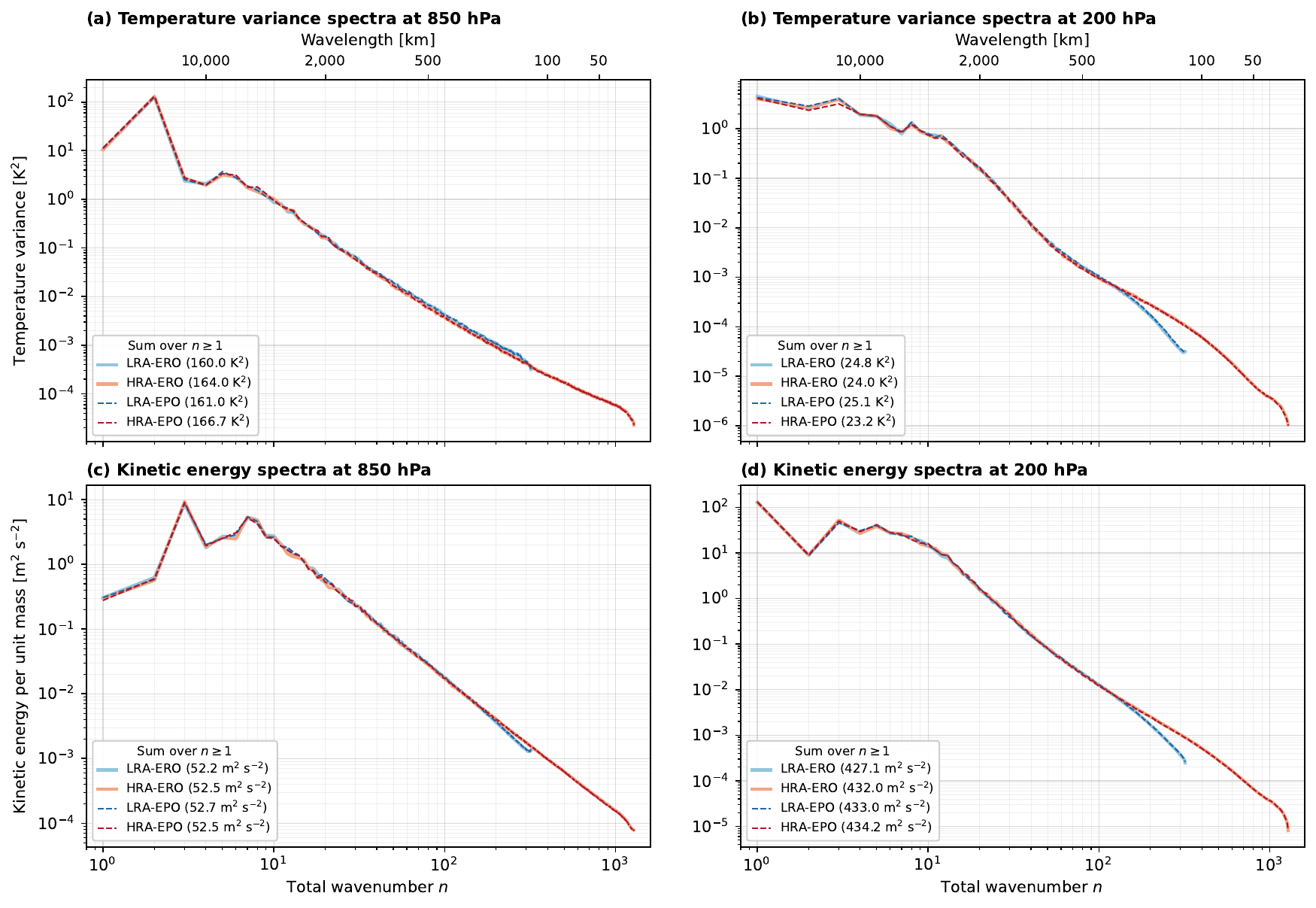}
    \centering
    \caption{Temperature variance (a, b) and kinetic energy per unit mass (c, d) spectra as a function of total wavenumber $n$ at 850 hPa (a, c) and 200 hPa (b, d) for day 32 of single-member eddy-permitting and eddy-rich reforecasts initialised on January 1st 1995 to 2016, averaged over all start dates. Values in parentheses give the spectra summed over $n \geq 1$, which correspond to the global variance and global mean kinetic energy, respectively. Wavelengths corresponding to each total wavenumber are shown along the top axes.}
    \label{fig:spectra}
\end{figure}

Figure \ref{fig:spectra} shows temperature variance and kinetic energy spectra at 850 and 200 hPa for coupled reforecasts initialised on January 1st. For a given atmospheric resolution, spectra from the eddy-rich and eddy-permitting configurations are almost indistinguishable at all resolved wavenumbers. The corresponding differences in global temperature variance and kinetic energy are within a few percent and are not statistically robust for any variable, level, or atmospheric resolution (table \ref{tab:all_scales_spectra_diff}). Similarly, we find no statistically robust differences when the analysis is restricted to wavelengths shorter than approximately 400 km ($n \geq 100$), where differences associated with the representation of the ocean mesoscale would be most likely to appear (table \ref{tab:small_scales_spectra_diff}). In contrast, the spectra are clearly sensitive to atmospheric resolution at wavelengths below $\sim$400 km, where the Tco319 configurations contain less variance and kinetic energy than their Tco1279 counterparts. We find similar results for reforecasts initialised on other start dates (not shown). These diagnostics characterise the impact on the free troposphere rather than the marine atmospheric boundary layer, where the local imprint of mesoscale SST anomalies is expected to be strongest (section \ref{section:air_sea_regressions}). Nevertheless, they demonstrate that increased ocean resolution does not meaningfully modify the distribution of tropospheric temperature variance or kinetic energy at any resolved scale in these experiments.

% -------------------------
% S T A R T   T A B L E S
% -------------------------
\begin{table}[!htbp]
\centering
\caption{Regional means of linear regression slopes (m s$^{-1}$ K$^{-1}$) of high-passed weekly mean 10 m wind speed anomalies on high-passed SST anomalies for days 26 to 32 of forecasts initialised in the extended boreal winter (ONDJFM) and summer (AMJJAS) between 2008 and 2016. Observation-based estimates use ESA-CCI SSTs and Copernicus Marine Service scatterometer-corrected winds. The Gulf Stream (37$^{\circ}$N-45$^{\circ}$N, 75$^{\circ}$W-40$^{\circ}$W), Southern Ocean (35$^{\circ}$S-55$^{\circ}$S), and Kuroshio (35$^{\circ}$N-42$^{\circ}$N, 140$^{\circ}$E-170$^{\circ}$E) regions are highlighted as boxes in figure \ref{fig:wind_sst_correlation}b.}
\label{tab:wind_sst_slopes}
\renewcommand{\arraystretch}{1.3}
\footnotesize
% Regional mean high-passed slope m s$^{-1}$ K$^{-1}$
% Start dates 2008-2016, days 26 to 32, boxcar_6x18 filter, configuration suffix '_cmems_wind'
% North Atlantic: latitude (30, 55), longitude (280, 340)
% Southern Ocean: latitude (-60, -40), longitude (0, 360)
% North Pacific: latitude (30, 50), longitude (140, 220)
\begin{tabular}{lcccccc}
\hline
 & \multicolumn{2}{c}{\textbf{Gulf Stream}} & \multicolumn{2}{c}{\textbf{Southern Ocean}} & \multicolumn{2}{c}{\textbf{Kuroshio}} \\
\textbf{Dataset} & \textbf{ONDJFM} & \textbf{AMJJAS} & \textbf{ONDJFM} & \textbf{AMJJAS} & \textbf{ONDJFM} & \textbf{AMJJAS} \\
\hline
Observations & 0.15 & 0.07 & 0.18 & 0.29 & 0.20 & 0.13 \\
LRA-EPO & 0.15 & 0.03 & 0.12 & 0.26 & 0.18 & 0.10 \\
LRA-ERO & 0.17 & 0.09 & 0.12 & 0.26 & 0.19 & 0.12 \\
HRA-EPO & 0.17 & 0.11 & 0.18 & 0.28 & 0.20 & 0.14 \\
HRA-ERO & 0.16 & 0.10 & 0.17 & 0.29 & 0.19 & 0.12 \\
\hline\hline
\end{tabular}
\end{table}

\begin{table}[!htbp]
\centering
\caption{Impact of ocean resolution on global temperature variance (T), kinetic energy (KE), and the divergent (KE$_{\mathrm{div}}$) and rotational (KE$_{\mathrm{rot}}$) components of kinetic energy at 200 and 850 hPa. Differences are calculated as eddy-rich minus eddy-permitting values, expressed as a percentage of the eddy-permitting values, for day 32 of single-member reforecasts initialised on January 1st 1995 to 2016. Values are derived from the spectra shown in figure \ref{fig:spectra} summed over all total wavenumbers $n \geq 1$. Square brackets give 95\% confidence intervals estimated by bootstrap resampling of start dates, and differences with confidence intervals that exclude zero would be shown in bold.}
\label{tab:all_scales_spectra_diff}
\renewcommand{\arraystretch}{1.3}
\footnotesize
% Start dates 20170101
% Wavenumbers 1 <= n <= inf, 95% CI
\begin{tabular}{lcccccc}
\hline
\textbf{\boldmath Atmosphere} & \textbf{\boldmath Level (hPa)} & \textbf{\boldmath Lead (days)} & \textbf{\boldmath $T$ (\%)} & \textbf{\boldmath KE (\%)} & \textbf{\boldmath KE$_{\mathrm{div}}$ (\%)} & \textbf{\boldmath KE$_{\mathrm{rot}}$ (\%)} \\
\hline
Tco319 & 200 & 32 & $-$1.0 [$-$7.7, 6.6] & $-$1.4 [$-$3.7, 1.2] & 3.7 [$-$4.2, 12.8] & $-$1.5 [$-$3.8, 1.0] \\
Tco319 & 850 & 32 & $-$0.6 [$-$2.7, 1.7] & $-$1.0 [$-$4.3, 2.3] & 1.9 [$-$2.6, 6.6] & $-$1.2 [$-$4.4, 2.2] \\
Tco1279 & 200 & 32 & 3.6 [$-$1.5, 9.7] & $-$0.5 [$-$2.8, 1.7] & 2.1 [$-$5.3, 8.0] & $-$0.6 [$-$2.8, 1.6] \\
Tco1279 & 850 & 32 & $-$1.6 [$-$3.7, 0.6] & 0.1 [$-$3.0, 3.7] & 0.2 [$-$4.0, 4.2] & 0.1 [$-$3.2, 3.8] \\
\hline\hline
\end{tabular}
\end{table}

\begin{table}[!htbp]
\centering
\caption{As table \ref{tab:all_scales_spectra_diff}, but for spectra summed over total wavenumbers $n \geq 100$, which correspond to wavelengths shorter than approximately 400 km.}
\label{tab:small_scales_spectra_diff}
\renewcommand{\arraystretch}{1.3}
\footnotesize
% Start dates 20170101
% Wavenumbers 100 <= n <= inf, 95% CI
\begin{tabular}{lcccccc}
\hline
\textbf{\boldmath Atmosphere} & \textbf{\boldmath Level (hPa)} & \textbf{\boldmath Lead (days)} & \textbf{\boldmath $T$ (\%)} & \textbf{\boldmath KE (\%)} & \textbf{\boldmath KE$_{\mathrm{div}}$ (\%)} & \textbf{\boldmath KE$_{\mathrm{rot}}$ (\%)} \\
\hline
Tco319 & 200 & 32 & $-$4.5 [$-$10.6, 2.7] & $-$1.8 [$-$5.7, 2.5] & $-$1.1 [$-$5.2, 3.3] & $-$2.5 [$-$6.7, 2.3] \\
Tco319 & 850 & 32 & $-$1.2 [$-$4.7, 2.9] & $-$0.7 [$-$4.1, 2.3] & $-$2.7 [$-$5.8, 0.0] & 0.1 [$-$3.7, 3.5] \\
Tco1279 & 200 & 32 & 0.3 [$-$4.6, 5.9] & 2.0 [$-$1.2, 5.2] & 0.5 [$-$2.4, 3.7] & 3.6 [$-$0.6, 7.5] \\
Tco1279 & 850 & 32 & $-$0.1 [$-$3.1, 3.1] & $-$0.6 [$-$2.5, 1.6] & $-$0.2 [$-$2.0, 2.0] & $-$0.7 [$-$2.8, 1.7] \\
\hline\hline
\end{tabular}
\end{table}

% -------------------------
% E N D   T A B L E
% -------------------------

%==========================
% C O N C L U S I O N S
%==========================
\section{Discussion and conclusions}
\label{section:conclusions}

The oceans are the principal reservoir of heat in the climate system and interact with the atmosphere through exchanges of energy, moisture, and momentum that influence variability and predictability across all temporal and spatial scales. At longer timescales, the benefits of ocean-atmosphere coupling for initialised predictions and climate projections are well established. For example, seasonal-to-decadal forecast systems rely on the accurate representation of coupled ocean-atmosphere dynamics to capture predictable signals such as ENSO and its global teleconnections \citep{griffies1997predictability, stockdale1998global}. At shorter lead times, many studies have also highlighted the importance of air-sea coupling for the prediction of tropical atmospheric phenomena at timescales of days to weeks \citep{bender2007operational, woolnough2007role, seo2009evaluation, shelly2014coupled, brassington2015progress, mogensen2017tropical}. However, the benefits of ocean-atmosphere coupling for the extratropical atmosphere at medium-range and subseasonal lead times are more equivocal and likely sensitive to the relative balance between improvements associated with more realistic air-sea interactions and degradations arising from the development of biases at the coupling interface \citep{roberts2021hemispheric}. Here, we provide a systematic assessment of the role of ocean-atmosphere coupling on subseasonal predictability, including an exploration of the potential for additional improvements through increased horizontal resolution and a better representation of the ocean mesoscale.

We find the largest benefits of ocean-atmosphere coupling for subseasonal atmospheric predictability in the tropical troposphere, where robust and widespread improvements are evident from week one and become systematically stronger with increased forecast lead time (figures \ref{fig:mabs}-\ref{fig:coupling_impact_t200}). Tropical forecast improvements are also evident in the increased skill of MJO predictions (figure \ref{fig:mjo_corrs}), such that ensemble mean bivariate RMM correlations decay to a threshold value of 0.6 at day 26 in ten-member coupled reforecasts (ENS-C) compared to day 21 in ten-member uncoupled reforecasts (ENS-U). These results are generally consistent with previous studies and reflect the strong coupling between SST anomalies, turbulent air-sea fluxes, deep atmospheric convection, and the large-scale circulation in the tropics \citep{woolnough2007role, kim2010ocean, klingaman2014role, demott2015atmosphere}.

Despite the substantial improvements to extratropical SST and sea ice anomaly forecasts in ENS-C compared to ENS-U, we find that ocean-atmosphere coupling has a more limited impact on weekly mean extratropical atmospheric variability at subseasonal lead times (figures \ref{fig:mabs}-\ref{fig:coupling_impact_t200}). For example, the subseasonal forecast skill of daily mean NAO, PNA, and PVORTEX indices is consistent in ENS-C and ENS-U within our estimated 95\% confidence intervals for almost all lead times for all three circulation indices (figure \ref{fig:nao_etc}). Weekly mean forecasts of 200 hPa velocity potential and 200 hPa streamfunction anomalies are significantly improved with ocean-atmosphere coupling in the extratropics. However, these non-local fields are strongly influenced by the rotational and divergent winds in the tropics, and their improvements are not accompanied by similar improvements in local extratropical fields such as 200 hPa winds, 500 hPa geopotential height, or mean sea level pressure. These signals therefore most likely reflect improved tropical and subtropical skill rather than an improved representation of the extratropical circulation.

At climate timescales, an accurate representation of ocean eddies is crucial for a realistic simulation of ocean mass, heat, and salt transports and associated air-sea interactions and atmospheric feedbacks \citep{griffies2015impacts, hewitt2017will, hirschi2020atlantic}. However, at daily to weekly timescales, many of the known deficiencies of eddy-permitting models can be mitigated by accurate ocean initialisation \citep[e.g. ][]{roberts2022sensitivity}. Furthermore, depending on the details of the coupled model configuration, the effective SST resolution seen by the atmosphere may be more sensitive to the horizontal resolution of the atmosphere than the ocean (figure \ref{fig:gulf_stream}).

Here, we show that increasing ocean resolution from the eddy-permitting to eddy-rich regime in single-member subseasonal reforecasts results in small but statistically robust improvements to the climatological mean state of some ocean and sea ice fields, which are consistent across different atmospheric resolutions (figures \ref{fig:eddy_impact_tco319} and \ref{fig:eddy_impact_tco1279}). We also find some evidence for very small changes in the predictability of weekly mean ocean and sea ice anomalies (figures \ref{fig:eddy_impact_tco319} and \ref{fig:eddy_impact_tco1279}). However, the qualitative and quantitative details of these changes in deterministic skill are sensitive to atmospheric configuration and the choice of verification grid (figure \ref{fig:eddy_impact_sst_rmse}). 

In contrast, we do not find strong evidence that coupling to an eddy-rich ocean substantially improves weekly mean atmospheric forecast skill in single-member subseasonal reforecasts coupled to either Tco319 ($\Delta$x $\approx$ 35 km) or Tco1279 ($\Delta$x $\approx$ 9 km) configurations of the IFS atmosphere (figures \ref{fig:eddy_impact_tco319} and \ref{fig:eddy_impact_tco1279}). The differences in RMSE are generally small and not statistically significant across multiple variables and atmospheric resolutions. Nevertheless, we cannot rule out that more subtle improvements, below the detection threshold of these single-member experiments, could be identified with ensemble reforecasts and/or a larger sample of start dates. However, given our results in section \ref{section:results_coupling}, it is perhaps not surprising that small changes to extratropical SSTs associated with increased horizontal ocean resolution do not translate to substantial signals in the large-scale atmospheric circulation at subseasonal timescales. The main caveat to this conclusion is that it remains possible that biases in the representation of parameterised boundary layer processes or other aspects of the ocean-atmosphere coupling in the IFS coupled model could limit the sensitivity of our forecasts to the presence of mesoscale SST anomalies \citep{maloney2006assessment, bryan2010frontal}.

To investigate this possibility, we complement our score-based evaluation of atmospheric predictability with scale-dependent evaluation of air-sea interactions and atmospheric variability. All reforecast configurations reproduce the observed scale-dependence of the relationship between near-surface wind speed and SST, with negative correlations at large scales, where the atmosphere drives the ocean, and positive correlations at the ocean mesoscale, where the ocean drives the atmosphere (figure \ref{fig:wind_sst_correlation}). The strength of this mesoscale coupling is comparable to observation-based estimates over the extratropical oceans, but weaker than observed over the eastern equatorial Pacific (table \ref{tab:wind_sst_slopes}). It is also generally insensitive to model resolution, with the largest differences associated with atmospheric rather than ocean resolution over the Southern Ocean, which is consistent with the dependence of the effective SST resolution experienced by the atmosphere on the atmospheric grid (figure \ref{fig:gulf_stream}). 

Furthermore, increased ocean resolution does not significantly modify the spectra of temperature variance or kinetic energy in the free troposphere at any resolved scale. In contrast, there are detectable differences in spectra associated with changes in atmospheric horizontal resolution (figure \ref{fig:spectra}). These scale-dependent diagnostics are consistent with our evaluation of large-scale atmospheric forecast skill and indicate that increasing the horizontal resolution of the ocean from the eddy-permitting to eddy-rich regime does not necessarily translate to substantial impacts on the large-scale tropospheric variability at subseasonal lead times. These results may appear to contradict previous studies that have investigated the impact of ocean mesoscale features on atmospheric variability \citep[e.g.][]{minobe2008influence}. However, our eddy-permitting and eddy-rich configurations both use initial conditions derived from the same realistic eddy-rich ocean state estimate and therefore measure the additional value of higher ocean resolution during the coupled forecast. The impacts are thus limited by the rate at which the eddy-permitting and eddy-rich ocean states diverge over the 32-day trajectory. This is a very different experimental design to idealised studies that have identified large-scale atmospheric responses when realistic SST fronts and eddies are replaced by heavily smoothed boundary conditions. 

Our results are generally consistent with other studies that have found a limited atmospheric response to an improved representation of the ocean mesoscale at daily to weekly lead times \citep{roberts2022sensitivity, polichtchouk2025effects, reynolds2025impact}. Taken together, these studies have practical implications for the design of subseasonal coupled forecast systems. Based on the timing statistics of our reforecasts, the computational cost of the ocean component is approximately 18 to 27 times larger in our eddy-rich configurations than in the corresponding eddy-permitting configurations, depending on the atmospheric resolution and the allocated computational resources. These estimates are only indicative as they are based on statistics from a single start date and the job configurations were not fully optimised for computational efficiency at all resolution combinations. However, this considerable increase in cost does not lead to detectable improvements in large-scale atmospheric forecast skill or variability at subseasonal lead times in our single-member experiments. Our scale-dependent diagnostics also suggest that the mesoscale air-sea interactions experienced by the atmosphere can (in some cases) be more sensitive to atmospheric than ocean resolution (figures \ref{fig:gulf_stream} and \ref{fig:wind_sst_correlation}), such that any atmospheric benefits of an eddy-rich ocean may only emerge when it is coupled to a sufficiently high-resolution atmosphere. Nevertheless, other factors that have not been considered in this study may be important when considering the case for an eddy-rich ocean in a coupled forecast system. Such factors include the potential for more accurate regional and/or coastal ocean and sea ice forecast products, indirect impacts on forecast quality through improved initial conditions achieved with higher-resolution (coupled) data assimilation systems, and impacts on extreme events that are not captured in our deterministic evaluation of large-scale atmospheric circulation.

%==========================
% A C K N O W L E D G M E N T S 
%==========================

\section*{Acknowledgements}
The reforecast datasets used in this study are publicly available from the ECMWF data catalogue. Table \ref{tab:data} provides experiment identifiers and links to the relevant datasets. Data from the ERA5 reanalysis are available from \url{https://www.ecmwf.int/en/forecasts/dataset/ecmwf-reanalysis-v5}. The GLORYS12v1 reanalysis is available from the Copernicus Marine Service at \url{https://doi.org/10.48670/moi-00021}. CP was supported by the European Union's Destination Earth Initiative and relates to tasks entrusted by the European Union to the European Centre for Medium-Range Weather Forecasts implementing part of this initiative with funding by the European Union. The lead author acknowledges the use of generative AI tools (ChatGPT, OpenAI; Claude, Anthropic) to assist with drafting and debugging code, generating formatted \LaTeX{} tables, and to provide critical feedback on an early manuscript draft when prompted to act as a reviewer. These tools were used to suggest manuscript revisions to improve clarity, but were not used for study design, data analysis, or interpretation of results. All outputs were reviewed, validated, and edited by the lead author, who takes full responsibility for the content of the publication. 

\section*{Conflict of interest}
The authors declare no conflict of interest.

%==========================
% A P P E N D I X
%==========================
\appendixpage
\begin{appendices}
\renewcommand{\thetable}{\thesection\arabic{table}}
\setcounter{table}{0}

\section{Ocean and sea ice initial conditions derived from GLORYS12v1}
Ocean and sea ice initial conditions for the eddy-permitting ocean (EPO) and eddy-rich ocean (ERO) configurations described in section \ref{section:methods} are derived from forced NEMO4-SI$^3$ simulations that are tightly constrained to the eddy-rich GLORYS12v1 reanalysis \citep{jean2021copernicus}. This is an updated version of the method described by \citet{roberts2022sensitivity} and \citet{Pelletier2023EGU23}, which is employed to obtain ocean and sea ice initial conditions that are consistent across different horizontal resolutions. For both EPO and ERO configurations, we run ocean/sea-ice simulations with surface boundary conditions constrained by hourly data from ERA5 \citep{hersbach2020era5}. In addition, 3D ocean temperature, 3D ocean salinity, sea ice concentration, and sea ice thickness, are relaxed towards daily mean data from the GLORYS12v1 reanalysis using a simple Newtonian scheme. Ocean temperature and salinity are constrained throughout the full vertical water column with a 5-day relaxation timescale. Sea ice concentration and thickness are relaxed with a 10-day relaxation timescale, with increments distributed from the GLORYS12v1 grid-box average across five SI3 sea ice thickness categories using a Gamma function. For the EPO experiments,  GLORYS12v1 data are remapped conservatively to the lower-resolution eORCA025 model grid. The ERO experiments have a comparable horizontal resolution to the original GLORYS12v1 data and interpolation between grids is based on a bilinear approach. For both ERO and EPO configurations, we linearly interpolate between the 50 vertical levels of GLORYS12v1 and the 75 vertical levels used by the NEMO4-SI3 configurations described in this study. The restart files from these pseudo-analyses are then used to provide ocean and sea ice initial conditions for coupled experiments that are tightly constrained to the GLORYS12v1 reanalysis and consistent across eddy-permitting and eddy-rich resolutions.

\section{Data accessibility}
\label{section:data}

\begin{table}[!htbp]
\centering
\caption{Data accessibility for reforecast experiments used in this study.}
\label{tab:data}
\renewcommand{\arraystretch}{1.3}
\footnotesize
\begin{tabular}{llp{5cm}p{4cm}}
\hline
\textbf{Experiment} & \textbf{ECMWF ID} & \textbf{URL} & \textbf{DOI} \\
\hline
\coupled{}   & impq & \url{https://apps.ecmwf.int/ifs-experiments/rd/impq/} & \href{https://doi.org/10.21957/2c00-rw28}{10.21957/2c00-rw28} \\
\uncoupled{} & imq1 & \url{https://apps.ecmwf.int/ifs-experiments/rd/imq1/} & \href{https://doi.org/10.21957/4z2b-e329}{10.21957/4z2b-e329} \\
\lraepo{}    & i7au & \url{https://apps.ecmwf.int/ifs-experiments/rd/i7au/} & \href{https://doi.org/10.21957/3qas-ev05}{10.21957/3qas-ev05} \\
\hraepo{}    & ii61 & \url{https://apps.ecmwf.int/ifs-experiments/rd/ii61/} & \href{https://doi.org/10.21957/h2x3-j680}{10.21957/h2x3-j680} \\
\lraero{}    & i7u3 & \url{https://apps.ecmwf.int/ifs-experiments/rd/i7u3/} & \href{https://doi.org/10.21957/7sqw-n588}{10.21957/7sqw-n588} \\
\hraero{}    & ii62 & \url{https://apps.ecmwf.int/ifs-experiments/rd/ii62/} & \href{https://doi.org/10.21957/bfqb-3521}{10.21957/bfqb-3521} \\
\hline\hline
\end{tabular}
\end{table}

\end{appendices}

%==========================
%       F I G U R E S
%==========================
\newpage

%==========================
% B I B L I O G R A P H Y 
%==========================
\clearpage
\newpage
\bibliographystyle{rss}
\bibliography{References}

@article{von2023heat,
  title={Heat stored in the {Earth} system 1960--2020: where does the energy go?},
  author={Von Schuckmann, Karina and Mini{\`e}re, Audrey and Gues, Flora and Cuesta-Valero, Francisco Jos{\'e} and Kirchengast, Gottfried and Adusumilli, Susheel and Straneo, Fiammetta and Ablain, Micha{\"e}l and Allan, Richard P and Barker, Paul M and others},
  journal={Earth System Science Data},
  volume={15},
  number={4},
  pages={1675--1709},
  year={2023},
  publisher={Copernicus GmbH}
}

@article{zuo2024ecmwf,
  title={{ECMWF}'s next ensemble reanalysis system for ocean and sea ice: {ORAS6}},
  author={Zuo, Hao and Balmaseda, Magdalena Alonso and de Boisseson, E and Browne, P and Chrust, Marcin and Keeley, S and Mogensen, K and Pelletier, C and de Rosnay, P and Takakura, T},
  journal={ECMWF Newsletter},
  volume={180},
  pages={30--36},
  year={2024}
}

@article{keeley2018dynamic,
  title={Dynamic sea ice in the {IFS}},
  author={Keeley, Sarah and Mogensen, Kristian},
  journal={ECMWF Newsletter},
  volume={156},
  pages={23--29},
  year={2018}
}

@article{eastwood2014algorithm,
  title={Algorithm theoretical basis document for the {OSI} {SAF} global reprocessed sea ice concentration product},
  author={Eastwood, STEINAR and Lavergne, THOMAS and Tonboe, RASMUS},
  journal={EUMETSAT Network Satellite Application Facilities},
  volume={28},
  year={2014}
}

@article{zuo2019ecmwf,
  title={The {ECMWF} operational ensemble reanalysis--analysis system for ocean and sea ice: a description of the system and assessment},
  author={Zuo, Hao and Balmaseda, Magdalena Alonso and Tietsche, Steffen and Mogensen, Kristian and Mayer, Michael},
  journal={Ocean science},
  volume={15},
  number={3},
  pages={779--808},
  year={2019},
  publisher={Copernicus Publications G{\"o}ttingen, Germany}
}

@article{trenberth2001estimates,
  title={Estimates of meridional atmosphere and ocean heat transports},
  author={Trenberth, Kevin E and Caron, Julie M},
  journal={Journal of Climate},
  volume={14},
  number={16},
  pages={3433--3443},
  year={2001}
}

@article{donlon2012operational,
  title={The operational sea surface temperature and sea ice analysis ({OSTIA}) system},
  author={Donlon, Craig J and Martin, Matthew and Stark, John and Roberts-Jones, Jonah and Fiedler, Emma and Wimmer, Werenfrid},
  journal={Remote sensing of Environment},
  volume={116},
  pages={140--158},
  year={2012},
  publisher={Elsevier}
}

@article{uppala2005era,
  title={The {ERA}-40 re-analysis},
  author={Uppala, Sakari M and K{\aa}llberg, PW and Simmons, Adrian J and Andrae, U and Bechtold, V Da Costa and Fiorino, M and Gibson, JK and Haseler, J and Hernandez, A and Kelly, GA and others},
  journal={Quarterly Journal of the Royal Meteorological Society: A journal of the atmospheric sciences, applied meteorology and physical oceanography},
  volume={131},
  number={612},
  pages={2961--3012},
  year={2005},
  publisher={Wiley Online Library}
}

@article{lang2021more,
  title={More accuracy with less precision},
  author={Lang, Simon TK and Dawson, Andrew and Diamantakis, Michail and Dueben, Peter and Hatfield, Samuel and Leutbecher, Martin and Palmer, Tim and Prates, Fernando and Roberts, Christopher D and Sandu, Irina and others},
  journal={Quarterly Journal of the Royal Meteorological Society},
  volume={147},
  number={741},
  pages={4358--4370},
  year={2021},
  publisher={Wiley Online Library}
}

@article{ferrari2009ocean,
  title={Ocean circulation kinetic energy: reservoirs, sources, and sinks},
  author={Ferrari, Raffaele and Wunsch, Carl},
  journal={Annual Review of Fluid Mechanics},
  volume={41},
  number={1},
  pages={253--282},
  year={2009},
  publisher={Annual Reviews}
}

@article{karlowska2024two,
  title={Two-way feedback between the {Madden--Julian Oscillation} and diurnal warm layers in a coupled ocean--atmosphere model},
  author={Karlowska, Eliza and Matthews, Adrian J and Webber, Benjamin GM and Graham, Tim and Xavier, Prince},
  journal={Quarterly Journal of the Royal Meteorological Society},
  volume={150},
  number={764},
  pages={4113--4132},
  year={2024},
  publisher={Wiley Online Library}
}

@article{talley2013closure,
  title={Closure of the global overturning circulation through the {Indian}, {Pacific}, and {Southern} {Oceans}: schematics and transports},
  author={Talley, Lynne D},
  journal={Oceanography},
  volume={26},
  number={1},
  pages={80--97},
  year={2013},
  publisher={JSTOR}
}

@article{gregory2008transient,
  title={Transient climate response estimated from radiative forcing and observed temperature change},
  author={Gregory, Jonathan M and Forster, PM},
  journal={Journal of Geophysical Research: Atmospheres},
  volume={113},
  number={D23},
  year={2008},
  publisher={Wiley Online Library}
}

@article{kuhlbrodt2012ocean,
  title={Ocean heat uptake and its consequences for the magnitude of sea level rise and climate change},
  author={Kuhlbrodt, Till and Gregory, JM},
  journal={Geophysical Research Letters},
  volume={39},
  number={18},
  year={2012},
  publisher={Wiley Online Library}
}

@article{levitus2012world,
  title={World ocean heat content and thermosteric sea level change (0--2000 m), 1955--2010},
  author={Levitus, Sydney and Antonov, John I and Boyer, Tim P and Baranova, Olga K and Garcia, Hernan Eduardo and Locarnini, Ricardo Alejandro and Mishonov, Alexey V and Reagan, James R and Seidov, Dan and Yarosh, Evgeney S and others},
  journal={Geophysical Research Letters},
  volume={39},
  number={10},
  year={2012},
  publisher={Wiley Online Library}
}

@article{cassou2018decadal,
  title={Decadal climate variability and predictability: challenges and opportunities},
  author={Cassou, Christophe and Kushnir, Yochanan and Hawkins, Ed and Pirani, Anna and Kucharski, Fred and Kang, In-Sik and Caltabiano, Nico},
  journal={Bulletin of the American Meteorological Society},
  volume={99},
  number={3},
  pages={479--490},
  year={2018}
}

@article{robson2012causes,
  title={Causes of the rapid warming of the {North Atlantic Ocean} in the mid-1990s},
  author={Robson, Jon and Sutton, Rowan and Lohmann, Katja and Smith, Doug and Palmer, Matthew D},
  journal={Journal of Climate},
  volume={25},
  number={12},
  pages={4116--4134},
  year={2012}
}

@article{england2014recent,
  title={Recent intensification of wind-driven circulation in the {Pacific} and the ongoing warming hiatus},
  author={England, Matthew H and McGregor, Shayne and Spence, Paul and Meehl, Gerald A and Timmermann, Axel and Cai, Wenju and Gupta, Alex Sen and McPhaden, Michael J and Purich, Ariaan and Santoso, Agus},
  journal={Nature climate change},
  volume={4},
  number={3},
  pages={222--227},
  year={2014},
  publisher={Nature Publishing Group UK London}
}

@article{mcphaden1999genesis,
  title={Genesis and evolution of the 1997-98 {El Ni{\~n}o}},
  author={McPhaden, Michael J},
  journal={Science},
  volume={283},
  number={5404},
  pages={950--954},
  year={1999},
  publisher={American Association for the Advancement of Science}
}

@article{robson2012initialized,
  title={Initialized decadal predictions of the rapid warming of the {North Atlantic Ocean} in the mid 1990s},
  author={Robson, JI and Sutton, RT and Smith, DM},
  journal={Geophysical Research Letters},
  volume={39},
  number={19},
  year={2012},
  publisher={Wiley Online Library}
}

@article{neelin1998enso,
  title={{ENSO} theory},
  author={Neelin, J David and Battisti, David S and Hirst, Anthony C and Jin, Fei-Fei and Wakata, Yoshinobu and Yamagata, Toshio and Zebiak, Stephen E},
  journal={Journal of Geophysical Research: Oceans},
  volume={103},
  number={C7},
  pages={14261--14290},
  year={1998},
  publisher={Wiley Online Library}
}

@article{hoskins1981steady,
  title={The steady linear response of a spherical atmosphere to thermal and orographic forcing},
  author={Hoskins, Brian J and Karoly, David J},
  journal={Journal of Atmospheric Sciences},
  volume={38},
  number={6},
  pages={1179--1196},
  year={1981}
}

@article{trenberth1998progress,
  title={Progress during {TOGA} in understanding and modeling global teleconnections associated with tropical sea surface temperatures},
  author={Trenberth, Kevin E and Branstator, Grant W and Karoly, David and Kumar, Arun and Lau, Ngar-Cheung and Ropelewski, Chester},
  journal={Journal of Geophysical Research: Oceans},
  volume={103},
  number={C7},
  pages={14291--14324},
  year={1998},
  publisher={Wiley Online Library}
}

@article{alexander2002atmospheric,
  title={The atmospheric bridge: the influence of {ENSO} teleconnections on air--sea interaction over the global oceans},
  author={Alexander, Michael A and Blad{\'e}, Ileana and Newman, Matthew and Lanzante, John R and Lau, Ngar-Cheung and Scott, James D},
  journal={Journal of climate},
  volume={15},
  number={16},
  pages={2205--2231},
  year={2002}
}

@article{griffies1997predictability,
  title={Predictability of {North Atlantic} multidecadal climate variability},
  author={Griffies, Stephen M and Bryan, Kirk},
  journal={Science},
  volume={275},
  number={5297},
  pages={181--184},
  year={1997},
  publisher={American Association for the Advancement of Science}
}

@article{stockdale1998global,
  title={Global seasonal rainfall forecasts using a coupled ocean--atmosphere model},
  author={Stockdale, Tim N and Anderson, David LT and Alves, J Oscar S and Balmaseda, Magdalena A},
  journal={Nature},
  volume={392},
  number={6674},
  pages={370--373},
  year={1998},
  publisher={Nature Publishing Group UK London}
}

@article{brassington2015progress,
  title={Progress and challenges in short-to medium-range coupled prediction},
  author={Brassington, GB and Martin, MJ and Tolman, HL and Akella, Santha and Balmeseda, M and Chambers, CRS and Chassignet, E and Cummings, JA and Drillet, Y and Jansen, PAEM and others},
  journal={Journal of Operational Oceanography},
  volume={8},
  number={sup2},
  pages={s239--s258},
  year={2015},
  publisher={Taylor \& Francis}
}

@article{woolnough2007role,
  title={The role of the ocean in the {Madden--Julian Oscillation}: implications for {MJO} prediction},
  author={Woolnough, SJ and Vitart, F and Balmaseda, MA},
  journal={Quarterly Journal of the Royal Meteorological Society: A journal of the atmospheric sciences, applied meteorology and physical oceanography},
  volume={133},
  number={622},
  pages={117--128},
  year={2007},
  publisher={Wiley Online Library}
}

@article{roberts2016drivers,
  title={On the drivers and predictability of seasonal-to-interannual variations in regional sea level},
  author={Roberts, CD and Calvert, D and Dunstone, N and Hermanson, L and Palmer, MD and Smith, D},
  journal={Journal of Climate},
  volume={29},
  number={21},
  pages={7565--7585},
  year={2016}
}

@article{demott2015atmosphere,
  title={Atmosphere-ocean coupled processes in the {Madden-Julian} oscillation},
  author={DeMott, Charlotte A and Klingaman, Nicholas P and Woolnough, Steven J},
  journal={Reviews of Geophysics},
  volume={53},
  number={4},
  pages={1099--1154},
  year={2015},
  publisher={Wiley Online Library}
}

@article{bender2007operational,
  title={The operational {GFDL} coupled hurricane--ocean prediction system and a summary of its performance},
  author={Bender, Morris A and Ginis, Isaac and Tuleya, Robert and Thomas, Biju and Marchok, Timothy},
  journal={Monthly Weather Review},
  volume={135},
  number={12},
  pages={3965--3989},
  year={2007}
}

@article{mogensen2017tropical,
  title={Tropical cyclone sensitivity to ocean coupling in the {ECMWF} coupled model},
  author={Mogensen, Kristian S and Magnusson, Linus and Bidlot, Jean-Raymond},
  journal={Journal of Geophysical Research: Oceans},
  volume={122},
  number={5},
  pages={4392--4412},
  year={2017},
  publisher={Wiley Online Library}
}

@article{madden1972description,
  title={Description of global-scale circulation cells in the tropics with a 40--50 day period},
  author={Madden, Roland A and Julian, Paul R},
  journal={Journal of Atmospheric Sciences},
  volume={29},
  number={6},
  pages={1109--1123},
  year={1972}
}

@article{cassou2008intraseasonal,
  title={Intraseasonal interaction between the {Madden--Julian} oscillation and the {North Atlantic Oscillation}},
  author={Cassou, Christophe},
  journal={Nature},
  volume={455},
  number={7212},
  pages={523--527},
  year={2008},
  publisher={Nature Publishing Group UK London}
}

@article{lin2009observed,
  title={An observed connection between the {North Atlantic Oscillation} and the {Madden--Julian} oscillation},
  author={Lin, Hai and Brunet, Gilbert and Derome, Jacques},
  journal={Journal of Climate},
  volume={22},
  number={2},
  pages={364--380},
  year={2009}
}

@article{hall2001modulation,
  title={The modulation of tropical cyclone activity in the {Australian} region by the {Madden--Julian} oscillation},
  author={Hall, Jonty D and Matthews, Adrian J and Karoly, David J},
  journal={Monthly weather review},
  volume={129},
  number={12},
  pages={2970--2982},
  year={2001}
}

@article{liu2022intraseasonal,
  title={Intraseasonal variability of global land monsoon precipitation and its recent trend},
  author={Liu, Fei and Wang, Bin and Ouyang, Yu and Wang, Hui and Qiao, Shaobo and Chen, Guosen and Dong, Wenjie},
  journal={npj Climate and Atmospheric Science},
  volume={5},
  number={1},
  pages={30},
  year={2022},
  publisher={Nature Publishing Group UK London}
}

@article{kemball2002simulation,
  title={Simulation of the intraseasonal oscillation in the {ECHAM}-4 model: the impact of coupling with an ocean model},
  author={Kemball-Cook, Susan and Wang, Bin and Fu, Xiouhua},
  journal={Journal of the atmospheric Sciences},
  volume={59},
  number={9},
  pages={1433--1453},
  year={2002}
}

@article{zhang2006simulations,
  title={Simulations of the {Madden--Julian} oscillation in four pairs of coupled and uncoupled global models},
  author={Zhang, Chidong and Dong, Min and Gualdi, Silvio and Hendon, Harry H and Maloney, Eric D and Marshall, Andrew and Sperber, Kenneth R and Wang, Wanqiu},
  journal={Climate Dynamics},
  volume={27},
  number={6},
  pages={573--592},
  year={2006},
  publisher={Springer}
}

@article{liess2004intraseasonal,
  title={The intraseasonal oscillation in {ECHAM}4 {Part I}: coupled to a comprehensive ocean model},
  author={Liess, Stefan and Bengtsson, Lennart and Arpe, Klaus},
  journal={Climate Dynamics},
  volume={22},
  number={6},
  pages={653--669},
  year={2004},
  publisher={Springer}
}

@article{sperber2004madden,
  title={{Madden-Julian} variability in {NCAR} {CAM}2.0 and {CCSM}2.0},
  author={Sperber, KR},
  journal={Climate Dynamics},
  volume={23},
  number={3},
  pages={259--278},
  year={2004},
  publisher={Springer}
}

@article{shelly2014coupled,
  title={Coupled versus uncoupled hindcast simulations of the {Madden-Julian Oscillation} in the {Year of Tropical Convection}},
  author={Shelly, Ann and Xavier, Prince and Copsey, Dan and Johns, Tim and Rodr{\'\i}guez, Jos{\'e} M and Milton, Sean and Klingaman, Nicholas},
  journal={Geophysical Research Letters},
  volume={41},
  number={15},
  pages={5670--5677},
  year={2014},
  publisher={Wiley Online Library}
}

@article{seo2009evaluation,
  title={Evaluation of {MJO} forecast skill from several statistical and dynamical forecast models},
  author={Seo, Kyong-Hwan and Wang, Wanqiu and Gottschalck, Jon and Zhang, Qin and Schemm, Jae-Kyung E and Higgins, Wayne R and Kumar, Arun},
  journal={Journal of Climate},
  volume={22},
  number={9},
  pages={2372--2388},
  year={2009}
}

@article{klingaman2014role,
  title={The role of air--sea coupling in the simulation of the {Madden--Julian} oscillation in the {Hadley Centre} model},
  author={Klingaman, NP and Woolnough, SJ},
  journal={Quarterly Journal of the Royal Meteorological Society},
  volume={140},
  number={684},
  pages={2272--2286},
  year={2014},
  publisher={Wiley Online Library}
}

@article{kim2010ocean,
  title={Ocean--atmosphere coupling and the boreal winter {MJO}},
  author={Kim, Hye-Mi and Hoyos, Carlos D and Webster, Peter J and Kang, In-Sik},
  journal={Climate dynamics},
  volume={35},
  number={5},
  pages={771--784},
  year={2010},
  publisher={Springer}
}

@incollection{balmaseda2026role,
  title={The role of the ocean in subseasonal-to-seasonal predictability and prediction},
  author={Balmaseda, Magdalena A and DeMott, Charlotte and Reynolds, Carolyn A and Roberts, Christopher D and Subramanian, Aneesh},
  booktitle={Sub-seasonal to Seasonal Prediction},
  pages={271--320},
  year={2026},
  publisher={Elsevier}
}

@article{small2008air,
  title={Air--sea interaction over ocean fronts and eddies},
  author={Small, RJ d and deSzoeke, Simon P and Xie, SP and O’neill, L and Seo, H and Song, Q and Cornillon, P and Spall, M and Minobe, S},
  journal={Dynamics of Atmospheres and Oceans},
  volume={45},
  number={3-4},
  pages={274--319},
  year={2008},
  publisher={Elsevier}
}

@article{bryan2010frontal,
  title={Frontal scale air--sea interaction in high-resolution coupled climate models},
  author={Bryan, Frank O and Tomas, Robert and Dennis, John M and Chelton, Dudley B and Loeb, Norman G and McClean, Julie L},
  journal={Journal of Climate},
  volume={23},
  number={23},
  pages={6277--6291},
  year={2010}
}

@article{hewitt2017will,
  title={Will high-resolution global ocean models benefit coupled predictions on short-range to climate timescales?},
  author={Hewitt, Helene T and Bell, Michael J and Chassignet, Eric P and Czaja, Arnaud and Ferreira, David and Griffies, Stephen M and Hyder, Pat and McClean, Julie L and New, Adrian L and Roberts, Malcolm J},
  journal={Ocean Modelling},
  volume={120},
  pages={120--136},
  year={2017},
  publisher={Elsevier}
}

@article{chassignet2021importance,
  title={On the importance of high-resolution in large-scale ocean models},
  author={Chassignet, Eric P and Xu, Xiaobiao},
  journal={Advances in Atmospheric Sciences},
  volume={38},
  number={10},
  pages={1621--1634},
  year={2021},
  publisher={Springer}
}

@article{lee2018impact,
  title={Impact of {Gulf Stream} {SST} biases on the global atmospheric circulation},
  author={Lee, Robert W and Woollings, Tim J and Hoskins, Brian J and Williams, Keith D and O’Reilly, Christopher H and Masato, Giacomo},
  journal={Climate Dynamics},
  volume={51},
  number={9},
  pages={3369--3387},
  year={2018},
  publisher={Springer}
}

@article{roberts2021hemispheric,
  title={Hemispheric impact of {North Atlantic} {SST}s in subseasonal forecasts},
  author={Roberts, CD and Vitart, F and Balmaseda, MA},
  journal={Geophysical Research Letters},
  volume={48},
  number={4},
  pages={e2020GL0911446},
  year={2021},
  publisher={Wiley Online Library}
}

@article{hirschi2020atlantic,
  title={The {Atlantic} meridional overturning circulation in high-resolution models},
  author={Hirschi, Jo{\"e}l J-M and Barnier, Bernard and B{\"o}ning, Claus and Biastoch, Arne and Blaker, Adam T and Coward, Andrew and Danilov, Sergey and Drijfhout, Sybren and Getzlaff, Klaus and Griffies, Stephen M and others},
  journal={Journal of Geophysical Research: Oceans},
  volume={125},
  number={4},
  pages={e2019JC015522},
  year={2020},
  publisher={Wiley Online Library}
}

@article{roberts2022sensitivity,
  title={Sensitivity of {ECMWF} coupled forecasts to improved initialization of the ocean mesoscale},
  author={Roberts, Christopher D and Balmaseda, Magdalena A and Tietsche, Steffen and Vitart, Frederic},
  journal={Quarterly Journal of the Royal Meteorological Society},
  volume={148},
  number={749},
  pages={3694--3714},
  year={2022},
  publisher={Wiley Online Library}
}

@article{roberts2020time,
  title={The time-scale-dependent response of the wintertime {North Atlantic} to increased ocean model resolution in a coupled forecast model},
  author={Roberts, CD and Vitart, F and Balmaseda, MA and Molteni, F},
  journal={Journal of Climate},
  volume={33},
  number={9},
  pages={3663--3689},
  year={2020}
}

@article{roberts2023euro,
  title={{Euro-Atlantic} weather regimes and their modulation by tropospheric and stratospheric teleconnection pathways in {ECMWF} reforecasts},
  author={Roberts, Christopher D and Balmaseda, Magdalena A and Ferranti, Laura and Vitart, Frederic},
  journal={Monthly Weather Review},
  volume={151},
  number={10},
  pages={2779--2799},
  year={2023},
  publisher={American Meteorological Society}
}

@article{griffies2015impacts,
  title={Impacts on ocean heat from transient mesoscale eddies in a hierarchy of climate models},
  author={Griffies, Stephen M and Winton, Michael and Anderson, Whit G and Benson, Rusty and Delworth, Thomas L and Dufour, Carolina O and Dunne, John P and Goddard, Paul and Morrison, Adele K and Rosati, Anthony and others},
  journal={Journal of Climate},
  volume={28},
  number={3},
  pages={952--977},
  year={2015}
}

@article{hallberg2013using,
  title={Using a resolution function to regulate parameterizations of oceanic mesoscale eddy effects},
  author={Hallberg, Robert},
  journal={Ocean Modelling},
  volume={72},
  pages={92--103},
  year={2013},
  publisher={Elsevier}
}

@article{reynolds2025impact,
  title={Impact of ocean resolution on {Navy} {ESPC} forecast skill},
  author={Reynolds, Carolyn A and Crawford, William and Thoppil, Prasad G and Rushley, Stephanie S and Janiga, Matthew A and McLay, Justin J and Barton, Neil P},
  journal={Weather and Forecasting},
  volume={40},
  number={8},
  pages={1463--1479},
  year={2025},
  publisher={American Meteorological Society}
}

@article{maloney2006assessment,
  title={An assessment of the sea surface temperature influence on surface wind stress in numerical weather prediction and climate models},
  author={Maloney, Eric D and Chelton, Dudley B},
  journal={Journal of climate},
  volume={19},
  number={12},
  pages={2743--2762},
  year={2006}
}

@article{zampieri2023machine,
  title={A machine learning correction model of the winter clear-sky temperature bias over the {Arctic} sea ice in atmospheric reanalyses},
  author={Zampieri, Lorenzo and Arduini, Gabriele and Holland, Marika and Keeley, Sarah PE and Mogensen, Kristian and Shupe, Matthew D and Tietsche, Steffen},
  journal={Monthly Weather Review},
  volume={151},
  number={6},
  pages={1443--1458},
  year={2023}
}

@inproceedings{Pelletier2023EGU23,
  author    = {Pelletier, Charles and Roberts, Christopher D. and Vitart, Frederic and Balmaseda, Magdalena A. and Mogensen, Kristian and Sandu, Irina},
  title     = {Generating ocean initial condition for coupled forecasts through nudged {NEMO} experiments},
  booktitle = {EGU General Assembly 2023},
  year      = {2023},
  address   = {Vienna, Austria},
  dates     = {23--28 Apr 2023},
  publisher = {European Geosciences Union},
  doi       = {10.5194/egusphere-egu23-11718},
  url       = {https://doi.org/10.5194/egusphere-egu23-11718},
  note      = {EGU23-11718}
}

@article{jung2008scale,
  title={Scale-dependent verification of ensemble forecasts},
  author={Jung, Thomas and Leutbecher, Martin},
  journal={Quarterly Journal of the Royal Meteorological Society: A journal of the atmospheric sciences, applied meteorology and physical oceanography},
  volume={134},
  number={633},
  pages={973--984},
  year={2008},
  publisher={Wiley Online Library}
}

@article{lang2023ifs,
  title={{IFS} upgrade brings many improvements and unifies medium-range resolutions},
  author={Lang, Simon and Rodwell, Mark and Schepers, Dinand},
  journal={ECMWF Newsletter},
  volume={176},
  pages={21--28},
  year={2023}
}

@article{lledo2023scale,
  title={Scale-dependent verification of precipitation and cloudiness at {ECMWF}},
  author={Lled\'{o}, Lloren\c{c} and Haiden, Thomas and Schroettle, Josef and Forbes, Richard},
  journal={ECMWF Newsletter},
  volume={174},
  pages={18-22},
  year={2023}
}

@article{arduini2019impact,
  title={Impact of a multi-layer snow scheme on near-surface weather forecasts},
  author={Arduini, Gabriele and Balsamo, Gianpaolo and Dutra, Emanuel and Day, Jonathan J and Sandu, Irina and Boussetta, Souhail and Haiden, Thomas},
  journal={Journal of Advances in Modeling Earth Systems},
  volume={11},
  number={12},
  pages={4687--4710},
  year={2019},
  publisher={Wiley Online Library}
}

@article{roberts2024ifs,
  title={{IFS} upgrade improves near-surface wind and temperature forecasts},
  author={Roberts, Christopher D and Ingleby, Bruce and Geer, Alan and Hólm, Elias and Janousek, Martin and Prates, Fernando and Rodwell, Mark},
  journal={ECMWF Newsletter},
  volume={181},
  pages={16--25},
  year={2024}
}

@misc{leutbecher2024improving,
  title={Improving the physical consistency of ensemble forecasts by using {SPP} in the {IFS}},
  author={Leutbecher, Martin and Lang, Simon and Lock, Sarah-Jane and Roberts, Christopher D and Tsiringakis, Aristofanis},
  journal={ECMWF Newsletter},
  volume={181},
  pages={26--31},
  year={2024}
}

@misc{polichtchouk2025upgrade,
  title={Upgrade to {IFS} cycle 50r1},
  author={Polichtchouk, Inna and Massart, Sebastien and Kipling, Zak},
  journal={ECMWF Newsletter},
  volume={185},
  pages={24--32},
  year={2025}
}

@article{keeley2024introduction,
  title={Introduction of a new ocean and sea-ice model based on {NEMO4-SI3}},
  author={Keeley, S and Mogensen, K and Bidlot, JR and Alonso-Balmaseda, M and Hatfield, S},
  journal={ECMWF Newsletter},
  volume={180},
  pages={24--29},
  year={2024}
}

@article{kanehama2022evaluation,
  title={Evaluation and optimizaton of orographic drag in the {IFS}},
  author={Kanehama, T and Sandu, I and Beljaars, A and van Niekerk, A and Wedi, N and Boussetta, S and Lang, S and Johnson, S and Magnusson, L},
  journal={ECMWF Technical Memoranda},
  volume={893},
  pages={31},
  year={2022}
}

@article{ollinaho2017towards,
  title={Towards process-level representation of model uncertainties: stochastically perturbed parametrizations in the {ECMWF} ensemble},
  author={Ollinaho, Pirkka and Lock, Sarah-Jane and Leutbecher, Martin and Bechtold, Peter and Beljaars, Anton and Bozzo, Alessio and Forbes, Richard M and Haiden, Thomas and Hogan, Robin J and Sandu, Irina},
  journal={Quarterly Journal of the Royal Meteorological Society},
  volume={143},
  number={702},
  pages={408--422},
  year={2017},
  publisher={Wiley Online Library}
}

@techreport{madec2019nemo,
  author={Madec, Gurvan and Bourdallé-Badie, Romain and Chanut, Jérôme and Clementi, Emanuela and Coward, Andrew and Ethé, Christian and Iovino, Doroteaciro and Lea, Dan and Lévy, Claire and Lovato, Tomas and Martin, Nicolas and Masson, Sébastien and Mocavero, Silvia and Rousset, Clément and Storkey, Dave and Vancoppenolle, Martin and Müeller, Simon and Nurser, George and Bell, Mike and Samson, Guillaume},
  title={{NEMO} ocean engine},
  year={2019},
  institution={Institut Pierre-Simon Laplace (IPSL)},
  volume       = {27},
  version      = {4.0},
  year         = {2019},
  publisher    = {Zenodo},
  doi          = {10.5281/zenodo.3878122},
  url          = {https://doi.org/10.5281/zenodo.3878122}
}

@techreport{vancoppenolle2023si3,
  author = {Vancoppenolle, Martin and Rousset, Clément and Blockley, Edward and Aksenov, Yevgeny and Feltham, Daniel and Fichefet, Thierry and Garric, Gilles and Guémas, Virginie and Iovino, Doroteaciro and Keeley, Sarah and Madec, Gurvan and Massonnet, François and Ridley, Jeff and Schroeder, David and Tietsche, Steffen},
  title = {{SI3}},
  year         = {2023},
  publisher    = {Zenodo},
  doi          = {10.5281/zenodo.7534900},
  url          = {https://doi.org/10.5281/zenodo.7534900}
}

@article{leutbecher2017stochastic,
  title={Stochastic representations of model uncertainties at {ECMWF}: state of the art and future vision},
  author={Leutbecher, Martin and Lock, Sarah-Jane and Ollinaho, Pirkka and Lang, Simon TK and Balsamo, Gianpaolo and Bechtold, Peter and Bonavita, Massimo and Christensen, Hannah M and Diamantakis, Michail and Dutra, Emanuel and others},
  journal={Quarterly Journal of the Royal Meteorological Society},
  volume={143},
  number={707},
  pages={2315--2339},
  year={2017},
  publisher={Wiley Online Library}
}

@article{lock2019treatment,
  title={{Treatment of model uncertainty from radiation by the Stochastically Perturbed Parametrization Tendencies (SPPT) scheme and associated revisions in the ECMWF ensembles}},
  author={Lock, Sarah-Jane and Lang, Simon TK and Leutbecher, Martin and Hogan, Robin J and Vitart, Frederic},
  journal={Quarterly Journal of the Royal Meteorological Society},
  volume={145},
  pages={75--89},
  year={2019},
  publisher={Wiley Online Library}
}

@article{hersbach2020era5,
  title={{The ERA5 global reanalysis}},
  author={Hersbach, Hans and Bell, Bill and Berrisford, Paul and Hirahara, Shoji and Hor{\'a}nyi, Andr{\'a}s and Mu{\~n}oz-Sabater, Joaqu{\'\i}n and Nicolas, Julien and Peubey, Carole and Radu, Raluca and Schepers, Dinand and others},
  journal={Quarterly Journal of the Royal Meteorological Society},
  volume={146},
  number={730},
  pages={1999--2049},
  year={2020},
  publisher={Wiley Online Library}
}

@article{jean2021copernicus,
  title={The {Copernicus} global 1/12 oceanic and sea ice {GLORYS12} reanalysis},
  author={Lellouche, Jean-Michel and Greiner, Eric and Bourdall{\'e}-Badie, Romain and Garric, Gilles and Melet, Ang{\'e}lique and Dr{\'e}villon, Marie and Bricaud, Cl{\'e}ment and Hamon, Mathieu and Le Galloudec, Olivier and Regnier, Charly and others},
  journal={Frontiers in Earth Science},
  volume={9},
  pages={698876},
  year={2021},
  publisher={Frontiers Media SA}
}

@article{madden1971detection,
  title={{Detection of a 40--50 day oscillation in the zonal wind in the tropical Pacific}},
  author={Madden, Roland A and Julian, Paul R},
  journal={Journal of Atmospheric Sciences},
  volume={28},
  number={5},
  pages={702--708},
  year={1971}
}

@article{vitart2017madden,
  title={{Madden—Julian Oscillation prediction and teleconnections in the S2S database}},
  author={Vitart, Fr{\'e}d{\'e}ric},
  journal={Quarterly Journal of the Royal Meteorological Society},
  volume={143},
  number={706},
  pages={2210--2220},
  year={2017},
  publisher={Wiley Online Library}
}

@article{gottschalck2010framework,
  title={{A framework for assessing operational model MJO forecasts: a project of the CLIVAR Madden-Julian oscillation working group}},
  author={Gottschalck, Jon and Wheeler, M and Weickmann, K and Vitart, F and Savage, N and Lin, H and Hendon, H and Waliser, D and Sperber, K and Prestrelo, C and others},
  journal={Bull Am Meteorol Soc},
  volume={91},
  number={8},
  pages={1247--1258},
  year={2010}
}

@article{wheeler2004all,
  title={{An all-season real-time multivariate MJO index: Development of an index for monitoring and prediction}},
  author={Wheeler, Matthew C and Hendon, Harry H},
  journal={Monthly weather review},
  volume={132},
  number={8},
  pages={1917--1932},
  year={2004},
  publisher={American Meteorological Society}
}

@misc{sealevel_product,
  author = {{CMS}},
  title={{Global Ocean Gridded L 4 Sea Surface Heights And Derived Variables Reprocessed Copernicus Climate Service}}, 
  published={Copernicus Marine Service },
  howpublished={\url{https://data.marine.copernicus.eu/product/SEALEVEL_GLO_PHY_CLIMATE_L4_MY_008_057}},
  year = {2026}, 
  note = "[Online; accessed 2026-03-02]"
}

@article{merchant2019satellite,
  title={Satellite-based time-series of sea-surface temperature since 1981 for climate applications},
  author={Merchant, Christopher J and Embury, Owen and Bulgin, Claire E and Block, Thomas and Corlett, Gary K and Fiedler, Emma and Good, Simon A and Mittaz, Jonathan and Rayner, Nick A and Berry, David and others},
  journal={Scientific data},
  volume={6},
  number={1},
  pages={1--18},
  year={2019},
  publisher={Nature Publishing Group}
}

@article{good2020current,
  title={The current configuration of the {OSTIA} system for operational production of foundation sea surface temperature and ice concentration analyses},
  author={Good, Simon and Fiedler, Emma and Mao, Chongyuan and Martin, Matthew J and Maycock, Adam and Reid, Rebecca and Roberts-Jones, Jonah and Searle, Toby and Waters, Jennifer and While, James and others},
  journal={Remote Sensing},
  volume={12},
  number={4},
  pages={720},
  year={2020},
  publisher={Multidisciplinary Digital Publishing Institute}
}

@article{lavergne2019version,
  title={Version 2 of the {EUMETSAT} {OSI} {SAF} and {ESA} {CCI} sea-ice concentration climate data records},
  author={Lavergne, Thomas and S{\o}rensen, Atle Macdonald and Kern, Stefan and Tonboe, Rasmus and Notz, Dirk and Aaboe, Signe and Bell, Louisa and Dybkj{\ae}r, Gorm and Eastwood, Steinar and Gabarro, Carolina and others},
  journal={The Cryosphere},
  volume={13},
  number={1},
  pages={49--78},
  year={2019},
  publisher={Copernicus Publications G{\"o}ttingen, Germany}
}

@misc{ifsdoc,
  author       = "{ECMWF}",
  title        = "{IFS documentation}",
  howpublished = "\url{https://www.ecmwf.int/en/publications/ifs-documentation}",
  year         = "2026",
  note         = "Accessed: 2026-02-19",
  organization = "European Centre for Medium-Range Weather Forecasts (ECMWF)"
}

@article{roberts2025unbiased,
  title={Unbiased calculation, evaluation, and calibration of ensemble forecast anomalies},
  author={Roberts, Christopher D and Leutbecher, Martin},
  journal={Quarterly Journal of the Royal Meteorological Society},
  volume={151},
  number={771},
  pages={e4993},
  year={2025},
  publisher={Wiley Online Library}
}

@article{roberts2025ensemble,
  title     = {Ensemble reliability and the signal-to-noise paradox in {ECMWF} subseasonal forecasts},
  author    = {Roberts, Christopher David and Vitart, Frederic},
  year      = {2026},
  eprint    = {2411.17694},
  archivePrefix = {arXiv},
  primaryClass  = {physics.ao-ph},
  url       = {https://arxiv.org/abs/2411.17694}
}

@article{leutbecher2019ensemble,
  title={Ensemble size: {H}ow suboptimal is less than infinity?},
  author={Leutbecher, Martin},
  journal={Quarterly Journal of the Royal Meteorological Society},
  volume={145},
  pages={107--128},
  year={2019},
  publisher={Wiley Online Library}
}

@article{ferro2014fair,
  title={Fair scores for ensemble forecasts},
  author={Ferro, CAT},
  journal={Quarterly Journal of the Royal Meteorological Society},
  volume={140},
  number={683},
  pages={1917--1923},
  year={2014},
  publisher={Wiley Online Library}
}

@article{ferro2008effect,
  title={On the effect of ensemble size on the discrete and continuous ranked probability scores},
  author={Ferro, Christopher AT and Richardson, David S and Weigel, Andreas P},
  journal={Meteorological Applications: A journal of forecasting, practical applications, training techniques and modelling},
  volume={15},
  number={1},
  pages={19--24},
  year={2008},
  publisher={Wiley Online Library}
}

@article{storkey2018uk,
  title={{UK} {G}lobal {O}cean {GO6} and {GO7}: {A} traceable hierarchy of model resolutions},
  author={Storkey, David and Blaker, Adam T and Mathiot, Pierre and Megann, Alex and Aksenov, Yevgeny and Blockley, Edward W and Calvert, Daley and Graham, Tim and Hewitt, Helene T and Hyder, Patrick and others},
  journal={Geoscientific Model Development},
  volume={11},
  number={8},
  pages={3187--3213},
  year={2018},
  publisher={Copernicus GmbH}
}

@article{polichtchouk2025effects,
  title={Effects of Atmosphere and Ocean Horizontal Model Resolution on Tropical Cyclone and Upper-Ocean Response Forecasts in Four Major Hurricanes},
  author={Polichtchouk, Inna and Mogensen, Kristian S and Sanabia, Elizabeth R and Jayne, Steven R and Magnusson, Linus and Densmore, Casey R and Hatfield, Sam and Hadade, Ioan and Wedi, Nils and Anantharaj, Valentine and others},
  journal={Monthly Weather Review},
  volume={153},
  number={11},
  pages={2257--2278},
  year={2025},
  publisher={American Meteorological Society}
}

@article{chelton2010coupled,
  title={Coupled ocean-atmosphere interaction at oceanic mesoscales},
  author={Chelton, Dudley B and Xie, Shang-Ping},
  journal={Oceanography},
  volume={23},
  number={4},
  pages={52--69},
  year={2010},
  doi={10.5670/oceanog.2010.05}
}

@misc{wind_product,
  author = {{CMS}},
  title={{Global Ocean Hourly Reprocessed Sea Surface Wind and Stress from Scatterometer and Model}},
  published={Copernicus Marine Service},
  howpublished={\url{https://data.marine.copernicus.eu/product/WIND_GLO_PHY_L4_MY_012_006}},
  year = {2026},
  note = "[Online; accessed 2026-10-01]"
}

@article{yoneyama2020years,
  title={Years of the {M}aritime {C}ontinent},
  author={Yoneyama, Kunio and Zhang, Chidong},
  journal={Geophysical Research Letters},
  volume={47},
  number={12},
  pages={e2020GL087182},
  year={2020},
  publisher={Wiley Online Library}
}

@article{jacobs2023adapting,
  title={Adapting constrained scales to observation resolution in ocean forecasts},
  author={Jacobs, Gregg and D'Addezio, Joseph and Bartels, Brent and DeHaan, Chris and Barron, Charlie and Carrier, Matthew and Shcherbina, Andrey and Dever, Mathieu},
  journal={Ocean Modelling},
  volume={186},
  pages={102252},
  year={2023},
  publisher={Elsevier}
}

@article{minobe2008influence,
  title={Influence of the {Gulf Stream} on the troposphere},
  author={Minobe, Shoshiro and Kuwano-Yoshida, Akira and Komori, Nobumasa and Xie, Shang-Ping and Small, Richard Justin},
  journal={Nature},
  volume={452},
  number={7184},
  pages={206--209},
  year={2008},
  doi={10.1038/nature06690}
}

\end{document}